\documentclass[twocolumn]{aastex631}
\usepackage{amsmath}
\usepackage{booktabs}

\newcommand{\teff}{\mbox{$T_{\mathrm{eff}}$}} 

\newcommand{\fbol}{$F_\mathrm{bol}$}
\newcommand{\lbol}{$L_\mathrm{bol}$}

\newcommand{\mic}{\mbox{$\mu$m}} 
\graphicspath{{./}{/}}

\begin{document}

\title{Precise Bolometric Luminosities and Effective Temperatures of 23 late-T and Y dwarfs Obtained with JWST}
\shorttitle{Precise \lbol's and \teff's of Ultracool Dwarfs}
\author[0000-0002-6721-1844]{Samuel A. Beiler}
\affiliation{Ritter Astrophysical Research Center, Department of Physics \& Astronomy,
University of Toledo, 2801 W. Bancroft St.,
Toledo, OH 43606, USA}

\author[0000-0001-7780-3352]{Michael C. Cushing}
\affiliation{Ritter Astrophysical Research Center, Department of Physics \& Astronomy,
University of Toledo, 2801 W. Bancroft St.,
Toledo, OH 43606, USA}

\author[0000-0002-6721-1844]{J. Davy Kirkpatrick}
\affiliation{IPAC, Mail Code 100-22, Caltech, 1200 E. California Boulevard, Pasadena, CA 91125, USA}

\author[0000-0003-4269-260X]{Adam C. Schneider}
\affiliation{United States Naval Observatory, Flagstaff Station, 10391 West Naval Observatory Road, Flagstaff, AZ 86005, USA}

\author[0000-0003-1622-1302]{Sagnick Mukherjee}
\affiliation{Department of Astronomy and Astrophysics, University of California, Santa Cruz, 1156 High Street, Santa Cruz, CA 95064, USA}

\author[0000-0002-5251-2943]{Mark S. Marley}
\affiliation{Lunar and Planetary Laboratory, University of Arizona, 1629 E. University Boulevard, Tucson, AZ 85721, USA}

\author[0000-0001-7519-1700]{Federico Marocco}
\affiliation{IPAC, Mail Code 100-22, Caltech, 1200 E. California Boulevard, Pasadena, CA 91125, USA}

\author[0000-0002-4424-4766]{Richard L. Smart}
\affiliation{Istituto Nazionale di Astrofisica, Osservatorio Astrofisico di Torino, Strada Osservatorio 20, I-10025 Pino Torinese, Italy}

\begin{abstract}
We present infrared spectral energy distributions of 23 late-type T and Y dwarfs obtained with the James Webb Space Telescope.  The spectral energy distributions consist of NIRSpec PRISM and MIRI LRS spectra covering the $\sim$1--12 $\mu$m wavelength range at $\lambda/ \Delta \lambda \approx 100$ and broadband photometry at 15, 18, and 21 $\mu$m.  The spectra exhibit absorption features common to these objects including H$_2$O, CH$_4$, CO, CO$_2$, and NH$_3$.  Interestingly, while the spectral morphology changes relatively smoothly with spectral type at $\lambda < 3$ $\mu$m and $\lambda > 8$ $\mu$m, it shows no clear trend in the 5 $\mu$m region where a large fraction of the flux emerges.  The broad wavelength coverage of the data enables us to compute the first accurate measurements of the bolometric fluxes of cool brown dwarfs.  Combining these bolometric fluxes with parallaxes from Spitzer and HST, we also obtain the first accurate bolometric luminosities of these cool dwarfs.  We then used the Sonora Bobcat solar metallicity evolutionary models to estimate the radii of the dwarfs which results in effective temperature estimates ranging from $\sim$1000 to 350 K with a median uncertainty of $\pm$20 K which is nearly an order of magnitude improvement over previous work. We also discuss how various portions of the spectra either do or do not exhibit a clear sequence when ordered by their effective temperatures.
\end{abstract}


\section{Introduction} \label{sec:intro}
Over the last two centuries, astronomers have advanced our understanding of the life of stars, largely through measuring fundamental stellar parameters such as mass ($M$), radius ($R$), and bolometric luminosity (\lbol), as well as derived parameters such as effective temperature (\teff=(\lbol/4$\pi\sigma R^2$)$^{1/4}$). A decades-long observational effort was needed to obtain these measurements for a broad population, with stellar parallaxes requiring precise astrometric observations \citep[e.g.][]{bessel_parallax_1838, campbell_stellar_1913}, bolometric fluxes requiring broad-wavelength spectral energy distributions \citep[e.g.][]{wilsing_effektive_1919,coblentz_effective_1922}, and radii and masses requiring time series photometry and spectroscopy of eclipsing spectroscopic binaries \citep[e.g.][]{frost_spectroscopy_1896,kuiper_empirical_1938}. Due to this effort, it is now possible to infer these properties for any star with a known spectral type \citep[e.g.][]{russell_relations_1914,payne_absorption_1924}. With this comes the ability to determine the mass function of stars, which is core to our understanding of star formation, galactic evolution, and even the early Universe \citep[e.g.][]{mckee_theory_2007,hopkins_dawes_2018,smith_evidence_2020, kirkpatrick_initial_2024}.  Large volume-limited samples of stars were painstakingly compiled in the latter half of the 20th century, resulting in a stellar field mass function that can be fit by a power law with a Salpeter slope at higher masses, and a shallower power law or log normal at lower masses \citep[e.g.][]{salpeter_luminosity_1955,bastian_universal_2010,kroupa_salpeter_2019}.

A similar advancement has been underway for brown dwarfs since the discovery of the  first brown dwarfs Gl 229B, Teide 1, and PPl 15 \citep{nakajima_discovery_1995,rebolo_discovery_1995, stauffer_ccd-based_1994, basri_lithium_1996}. However, progress over these 30 years has been hampered by observational and fundamental difficulties: 1) brown dwarfs do not obey a mass-luminosity relation like their stellar counterparts, as without an internal heat source they cool ``inexorably like dying embers plucked from a fire" \citep{burrows_non-gray_1997}, 2) the brown dwarf binary fraction is only 5$-$20\% compared to the $\sim$50\% for solar stars and higher for early spectral types \citep[e.g.][]{fontanive_constraining_2018}, making accurate radius and mass measurements difficult, and 3) brown dwarfs emit a significant portion of their radiation in the near- and mid-infrared ($>$1 \mic). This last difficulty worsens for later spectral types, with late-T and Y dwarfs (\teff$<$$\sim$1000 K) emitting over 80\% of their flux at wavelength redward of 3 \mic{}, and doing so faintly ($M_{[4.5]}>13$). This observational difficulty has been overcome with ground observations \citep[e.g.][]{skemer_first_2016,morley_l_2018, miles_observations_2020} and space telescopes like Spitzer Space Telescope \citep{werner_spitzer_2004}, WISE \citep{wright_wide-field_2010}, and AKARI \citep{murakami_infrared_2007}, but only for a handful of objects have sufficient spectral coverage for accurate bolometric fluxes and luminosities. This has made estimating the effective temperatures of these late objects difficult, resulting in uncertainties of order 20\% \citep{dupuy_distances_2013,kirkpatrick_field_2021}.

A lack of precise effective temperatures impacts our ability to construct a reliable mass function. Since brown dwarfs have no mass-luminosity relation, we cannot generate the mass function directly from the luminosities or effective temperatures of a field population. Instead, the most commonly used method is to construct a volume-limited luminosity (or effective temperature) distribution, and then fit this distribution with simulated populations created from an assumed mass function and birth rate, and evolved using evolutionary model tracks \citep{burgasser_t_2004,filippazzo_fundamental_2015,kirkpatrick_field_2021,best_volume-limited_2024, kirkpatrick_initial_2024}. From previous work collecting a large sample of M, L, and T dwarfs with precise luminosities and parallaxes, it has been shown that a power-law mass function with $\alpha \approx$ 0.6 with a minimum formation mass best fits local volume-limited samples \citep{kirkpatrick_field_2021,best_volume-limited_2024,kirkpatrick_initial_2024, raghu_simulating_2024}. Similar results have been found in young clusters \citep[e.g.][]{lodieu_wide_2007,ramirez_new_2012,alves_de_oliveira_low_2013}. However, due to the aforementioned lack of a large sample of late-T and Y dwarfs with reliable effective temperatures, the low end of this mass function is not well constrained. This includes limited constraints on the minimum formation mass. 

The James Webb Space Telescope \citep[JWST,][]{gardner_james_2023}, with its unprecedented sensitivity and broad spectral coverage, is a natural answer to our observational difficulties. For the late-T and Y dwarfs, JWST observations are capable of collecting $>$90\% of the flux, which along with precise parallaxes can lower the bolometric luminosity fractional uncertainty below 6\% \citep{beiler_first_2023}. These precise luminosities, along with theoretical radii often determined from evolutionary models, allow for the calculation of precise effective temperatures for objects with broad-wavelength JWST spectral energy distributions. With a representative sample of objects with these observations, it will be possible to improve the effective temperature estimates of hundreds of ultracool objects within the current volume-limited sample. This in turn will further our understanding of the substellar mass function and ultracool atmospheres.

In this paper we present the spectral energy distributions of 23 late-T and Y dwarfs, consisting of low resolution spectra from $\sim$1--12 \mic{} and photometry out to 21 \mic, and determine \lbol{} and \teff{} for the sample. \citet{beiler_first_2023} previously presented the spectral energy distribution of one of the objects from this sample (WISE J0359$-$54). In $\S$\ref{sec:sample}, we discuss our sample selection for this project, whose observations and reduction we report in $\S$\ref{sec:obvs} and $\S$\ref{sec:reduc}. These are the first near-infrared spectra for two of our objects, which we classify in $\S$\ref{sec:nirSpecType}. In $\S$\ref{sec:spec} we remark on the spectral morphology of these broad-wavelength spectral energy distributions and the large swath of molecular absorption features they contain. We then calculate \fbol, \lbol, and \teff{} for our whole sample in $\S$\ref{sec:lbolandteff}, and in $\S$\ref{sec:Varients} we explore how incomplete spectral energy distributions impact \teff and whether we can correct for their incompleteness. With effective temperatures in hand, we explore how absorption features change as a function of \teff{} in $\S$\ref{sec:specEvol}.

\section{Sample Selection} \label{sec:sample}
\begin{figure}
\includegraphics[width=.45\textwidth]{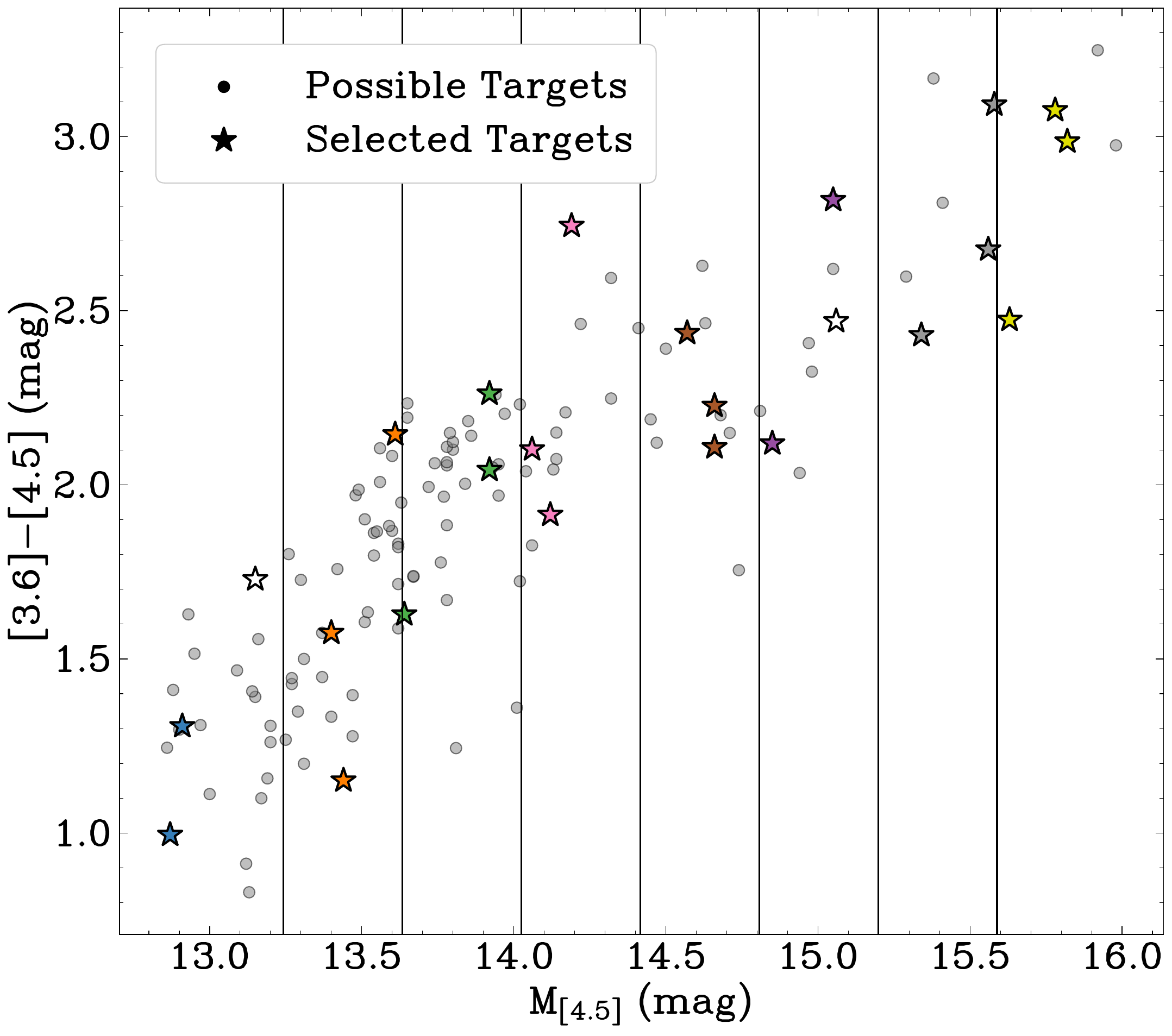}
\centering
\caption{The 143 potential targets (circles) from which we selected our sample (stars). The potential targets have spectral types of T6 or later, fall within 20 pc of the Sun, have fractional parallax uncertainties less than 5\%, and are not suspected of being binaries or subdwarfs. The potential targets were divided into 8 bins in }absolute [4.5] magnitude, which are color coded and marked with vertical lines. We selected 3 targets from each bin that span the [3.6]$-$[4.5] color space. The targets marked as white stars had parts of their observations fail. \label{fig:Sample}
\end{figure}

We selected our sample for this program (GO 2302, PI Cushing) from the 243 brown dwarfs with a spectral type of T6 or later within a 20 pc radius of the Sun \citep{kirkpatrick_preliminary_2019,kirkpatrick_field_2021}. We chose this spectral type cut-off so that there is overlap between our sample and that of \citet{filippazzo_fundamental_2015} who derived effective temperatures for 198 M, L, and T dwarfs. As much as we would like to observe all 243 of these brown dwarfs with JWST, such a proposal is untenable so we selected a representative sample. We eliminated suspected subdwarfs and binaries, and objects with a fractional parallax uncertainty greater than 5\%. This left us with 143 potential targets, which are shown in Figure~\ref{fig:Sample}. We placed these objects in eight bins based on their absolute [4.5] Spitzer magnitudes and selected three objects from each bin (shown as stars in the figure) based on their [3.6]$-$[4.5] colors: one with a roughly average color, one with a redder than average color, and one with a bluer than average color. The properties of these 24 objects are given in Table \ref{tbl:prop}. Hereafter we will abbreviate the full designations of these objects as HHMM+DD. 

\begin{deluxetable*}{ll|r|cc|cc|cccc|c}
\tablecaption{Astrometric and Photometric Properties of Our Sample\label{tbl:prop}}
\tabletypesize{\small}
\tablehead{
\colhead{Name} &
\colhead{Spec.} &
\colhead{$\varpi_\mathrm{abs}$} &
\colhead{IRAC [3.6]} &
\colhead{IRAC [4.5]} & 
\colhead{W1} &
\colhead{W2} & 
\colhead{F1000W}&
\colhead{F1500W}&
\colhead{F1800W}&
\colhead{F2100W}&
\colhead{Ref.$^\tablenotemark{b}$} \\
\colhead{} &
\colhead{type} &
\colhead{(mas)} &
\colhead{(mag)} &
\colhead{(mag)} &
\colhead{(mag)} &
\colhead{(mag)} &
\colhead{(mag)$^\tablenotemark{a}$} &
\colhead{(mag)$^\tablenotemark{a}$} &
\colhead{(mag)$^\tablenotemark{a}$} &
\colhead{(mag)$^\tablenotemark{a}$} &
\colhead{}
}
\startdata
\rotate
WISE J024714.52+372523.5       &T8   &$ 64.8\pm  2.0$ &$ 16.70\pm  0.04$ &$ 14.55\pm  0.02$ &$ 17.74\pm  0.09$ &$ 14.58\pm  0.02$ &$ 13.15$ &$ 12.64$ &$ 12.62$ &$ 12.63$ &14,14,3,5,9,1\\
WISEPA J031325.96+780744.2     &T8.5 &$135.6\pm  2.8$ &$ 15.31\pm  0.02$ &$ 13.27\pm  0.02$ &$ 15.98\pm  0.02$ &$ 13.22\pm  0.01$ &$ 11.62$ &$ 11.01$ &$ 10.92$ &$ 10.99$ &13,13,2,5,9,1\\
WISE J035934.06$-$540154.6       &Y0   &$ 73.6\pm  2.0$ &$ 17.55\pm  0.07$ &$ 15.33\pm  0.02$ &$ 19.21\pm  0.22$ &$ 15.41\pm  0.03$ &$ 13.66$ &$ 12.84$ &$ 12.65$ &$ 12.62$ &10,10,2,5,9,1\\
WISE J043052.92+463331.6       &T8   &$ 96.1\pm  2.9$ &$ 16.13\pm  0.03$ &$ 14.22\pm  0.02$ &$ 18.16\pm  0.17$ &$ 14.31\pm  0.02$ &$ 13.22$ &$ 12.51$ &$ 12.33$ &$ 12.40$ &14,14,2,5,9,1\\
WISE J053516.80$-$750024.9       &Y1   &$ 68.7\pm  2.0$ &$ 17.75\pm  0.08$ &$ 15.01\pm  0.02$ &$ 18.37\pm  0.10$ &$ 15.00\pm  0.02$ &$ 13.39$ &$ 12.30$ &$ 12.04$ &$ 11.95$ &10,10,2,5,9,1\\
WISE J073444.02$-$715744.0       &Y0   &$ 74.5\pm  1.7$ &$ 17.65\pm  0.08$ &$ 15.21\pm  0.02$ &$ 19.11\pm  0.21$ &$ 15.24\pm  0.02$ &$ 13.47$ &$ 12.63$ &$ 12.50$ &$ 12.51$ &10,10,2,5,9,1\\
WISE J082507.35+280548.5       &Y0.5 &$152.6\pm  2.0$ &$ 17.33\pm  0.06$ &$ 14.65\pm  0.02$ &$ 18.30\pm  0.18$ &$ 14.65\pm  0.02$ &$ 12.44$ &$ 11.29$ &$ 11.07$ &$ 10.97$ &11,11,2,5,9,1\\
ULAS J102940.52+093514.6       &T8   &$ 68.6\pm  1.7$ &$ 16.08\pm  0.03$ &$ 14.46\pm  0.02$ &$ 16.76\pm  0.04$ &$ 14.43\pm  0.02$ &$ 12.59$ &$ 12.25$ &$ 12.21$ &$ 12.33$ &18,19,3,5,9,1\\
CWISEP J104756.81+545741.6     &Y1   &$ 68.1\pm  4.9$ &$ 18.73\pm  0.17$ &$ 16.26\pm  0.02$ &$ 19.36$$^\tablenotemark{c}$&$ 16.18\pm  0.06$ &$ 14.15$ &$ 13.23$ &$ 13.10$ &$ 13.14$ &7,1,1,7,9,1\\
WISE J120604.38+840110.6       &Y0   &$ 84.7\pm  2.1$ &$ 17.34\pm  0.06$ &$ 15.22\pm  0.02$ &$ 18.09\pm  0.09$ &$ 15.17\pm  0.02$ &$ 12.91$ &$ 12.20$ &$ 11.99$ &$ 11.98$ &11,11,2,5,9,1\\
SDSSp J134646.45$-$003150.4      &T6.5 &$ 69.2\pm  2.3$ &$ 14.67\pm  0.02$ &$ 13.67\pm  0.02$ &$ 15.48\pm  0.02$ &$ 13.69\pm  0.01$ &$ 11.91$ &$ 11.71$ &$ 11.66$ &$ 11.65$ &20,21,4,5,9,1\\
WISEPC J140518.40+553421.4     &Y0.5 &$158.2\pm  2.6$ &$ 16.88\pm  0.05$ &$ 14.06\pm  0.02$ &$ 18.27\pm  0.13$ &$ 14.10\pm  0.01$ &$ 12.53$ &$ 11.40$ &$ 11.16$ &$ 11.09$ &12,11,2,5,9,1\\
CWISEP J144606.62$-$231717.8     &Y1   &$103.8\pm  5.0$ &$ 18.91\pm  0.04$ &$ 15.92\pm  0.02$ &$ 19.65$$^\tablenotemark{c}$&$ 15.96\pm  0.07$ &$ 13.67$ &$ 12.46$ &$ 12.22$ &$ 12.18$ &7,1,1,8,9,1\\
WISE J150115.92$-$400418.4       &T6   &$ 72.8\pm  2.3$ &$ 15.28\pm  0.02$ &$ 14.13\pm  0.02$ &$ 16.13\pm  0.03$ &$ 14.23\pm  0.02$ &$ 12.53$ &$ 12.32$ &$ 12.33$ &$ 12.29$ &16,15,2,5,9,1\\
WISEPA J154151.66$-$225025.2     &Y1   &$166.9\pm  2.0$ &$ 16.66\pm  0.04$ &$ 14.23\pm  0.02$ &$ 17.61\pm  0.17$ &$ 14.22\pm  0.03$ &$ 11.98$ &$ 10.93$ &$ 10.71$ &$ 10.66$ &12,11,2,5,9,1\\
SDSS J162414.37+002915.6       &T6   &$ 91.8\pm  1.2$ &$ 14.41\pm  0.02$ &$ 13.10\pm  0.02$ &$ 15.15\pm  0.02$ &$ 13.13\pm  0.01$ &$ 11.74$ &$ 11.53$ &$ 11.52$ &$ 11.58$ &22,21,4,5,9,1\\
WISEPA J195905.66$-$333833.7     &T8   &$ 83.9\pm  2.0$ &$ 15.36\pm  0.02$ &$ 13.79\pm  0.02$ &$ 16.14\pm  0.03$ &$ 13.80\pm  0.01$ &$ 11.96$ &$ 11.77$ &$ 11.54$ &$ 11.55$ &13,13,2,5,9,1\\
WISEPC J205628.90+145953.3     &Y0   &$140.8\pm  2.0$ &$ 16.03\pm  0.03$ &$ 13.92\pm  0.02$ &$ 17.50\pm  0.10$ &$ 13.93\pm  0.02$ &$ 11.74$ &$ 10.93$ &$ 10.79$ &$ 10.82$ &12,12,2,5,9,1\\
WISE J210200.15$-$442919.5       &T9   &$ 92.9\pm  1.9$ &$ 16.32\pm  0.04$ &$ 14.22\pm  0.02$ &$ 16.70\pm  0.04$ &$ 14.16\pm  0.02$ &$ 12.59$ &$ 11.98$ &$ 11.95$ &$ 11.99$ &10,10,6,5,9,1\\
WISEA J215949.54$-$480855.2      &T9   &$ 73.9\pm  2.6$ &$ 16.84\pm  0.05$ &$ 14.58\pm  0.02$ &$ 17.97\pm  0.12$ &$ 14.65\pm  0.02$ &$ 13.42$ &$ 12.72$ &$ 12.64$ &$ 12.67$ &15,15,2,5,9,1\\
WISE J220905.73+271143.9       &Y0   &$161.7\pm  2.0$ &$ 17.82\pm  0.09$ &$ 14.74\pm  0.02$ &$ 18.52\pm  0.20$ &$ 14.74\pm  0.02$ &$ 12.67$ &$ 11.48$ &$ 11.22$ &$ 11.12$ &13,17,2,5,9,1\\
WISEA J235402.79+024014.1      &Y1   &$130.6\pm  3.3$ &$ 18.11\pm  0.11$ &$ 15.01\pm  0.02$ &$ 18.05\pm  0.14$ &$ 15.02\pm  0.03$ &$ 13.44$ &$ 12.23$ &$ 11.99$ &$ 11.94$ &11,11,2,5,9,1\\
WISEPA J201824.96$-$742325.9      &T7   &$ 83.2\pm  1.9$ &$ 15.28\pm  0.02$ &$ 13.55\pm  0.02$ &$ 16.46\pm  0.04$ &$ 13.60\pm  0.01$ &$ 12.51$ &$ 12.04$ &$ 12.00$ &$ 12.02$ &13,13,5,5,9,1\\
WISEPA J041022.71+150248.5     &Y0   &$ 151.3\pm  2.0$ &$ 16.64\pm 0.04$ &$ 14.17\pm  0.02$ &$ 18.02\pm  0.17$ &$ 14.10\pm  0.02$ &$ 12.27$ &$ 11.35$ &$ 11.24$ &$ 11.21$ &12,12,2,5,9,1\\
\enddata
\tablenotetext{a}{We assume a 3\% or 0.03 mag uncertainty for all JWST magnitudes. The zero point fluxes for the MIRI photometry are 37.826, 16.760, 11.722, and 8.850 Jy, respectively.}
\tablenotetext{b}{References for discovery, spectral type, parallax, IRAC photometry, WISE photometry, and JWST photometry.}
\tablenotetext{c}{These magnitudes are 95\% confidence upper limits (see https://irsa.ipac.caltech.edu/data/WISE/CatWISE/gator\_docs/catwise\_colDescriptions.html)}
\tablerefs{(1) This Work, (2) \citet{kirkpatrick_field_2021}, (3) \citet{best_hawaii_2020}, (4) \citet{tinney_infrared_2003}, (5) \citet{kirkpatrick_preliminary_2019}, (6) \citet{tinney_luminosities_2014}, (7) \citet{meisner_expanding_2020}, (8) \citet{marocco_improved_2020}, (9) CatWISE2020 Catalog \citep{eisenhardt_catwise_2020}, (10) \citet{kirkpatrick_further_2012}, (11) \citet{schneider_hubble_2015}, (12) \citet{cushing_discovery_2011}, (13) \citet{kirkpatrick_first_2011}, (14) \citet{mace_study_2013}, (15) \citet{tinney_new_2018}, (16) \citet{tinney_wise_2012}, (17) \citet{cushing_ultracool_2014}, (18) \citet{burningham_76_2013}, (19) \citet{thompson_nearby_2013}, (20) \citet{tsvetanov_discovery_2000}, (21) \citet{burgasser_unified_2006}, (22) \citet{strauss_discovery_1999}}
\end{deluxetable*}

At the time of the proposal's submission, CWISEP J1446$-$23 and CWISEP J1047+54, two of our selected objects, did not meet our 5\% parallax cutoff, but were undergoing astrometric follow-up by some of the authors with the Hubble Space Telescope (HST) to reach this level. With the follow up now complete, we computed their parallaxes using data from WISE/NEOWISE, Spitzer, and HST. The procedure resembles closely the one described in \citet{kirkpatrick_field_2021}, but with a few key differences, which we briefly summarize here. We measure source positions in the HST drizzle images provided by MAST using CASU imcore, a source detection routine. The x,y pixel coordinates measured by imcore are converted to $\alpha,\delta$ using the drizzled image WCS, and then matched to Gaia DR3 using a 3\arcsec\ matching radius. The positions of matching Gaia sources are propagated from the Gaia epoch (2016.0) to the epoch of the HST observations ($\sim$2021) using their measured parallax and proper motion. We then derived a six-parameter transformation between the HST and the Gaia coordinates by projecting both coordinates onto a tangent plane whose tangent point is defined by the CRVAL1 and CRVAL2 FITS header keywords. We then solved for the six parameters of the transformation, which account for offsets, rotation, and scaling, using the IDL routine regress. The residuals of the transformation are added in quadrature to the measurement errors to compute the final coordinate uncertainties. More details on the full reduction and astrometric calibration of the HST data are given in Marocco et al. (in prep.). The parallax and proper motions were computed using the calibrated $\alpha,\delta$ values as described in \citet{kirkpatrick_preliminary_2019}. Comparing the new values with the preliminary measurements from \citet{kirkpatrick_preliminary_2019}, we see that the addition of just one HST measurement at the extreme of the previously poorly sampled parallactic ellipse allowed us to dramatically improve the quality of our fit. This results in fractional uncertainties of 4.8\% and 7.2\% for CWISEP J1446$-$23 and CWISEP J1047+54, respectively. These newly calculated parallaxes are given in Table 1.

\section{Observations} \label{sec:obvs}
We observed our sample using JWST's Near Infrared Spectrograph \citep[NIRSpec,][]{jakobsen_near-infrared_2022} and Mid Infrared Instrument \citep[MIRI, ][]{rieke_mid-infrared_2015}, collecting low-resolution spectroscopy with both instruments and broadband photometry with MIRI (GO 2302, PI Cushing). NIRSpec was used in fixed-slit mode with the CLEAR/PRISM filter and the S200A1 slit ($0\farcs2\times3\farcs2$) and SUBS200A1 subarray. This filter allowed us to obtain a spectrum from 0.6--5.3 \mic{} with a resolving power of $R = {\lambda}/{\Delta \lambda} = 30-300$. The observations were executed with the NRSRAPID readout pattern and a 5 point dither. MIRI spectroscopy was carried out using the low-resolution spectrometer (LRS) in slit mode ($0\farcs51\times4\farcs7$) and the FULL subarray. We used the FASTR1 readout mode and a 2-point ``along slit nod" dither. MIRI LRS allowed us to obtain a spectrum from $\sim$5--14 \mic{} with a resolving power of 50$-$200. We chose to truncate the red edge of the spectrum where the signal-to-noise ratio first falls below 10 past 12 \mic. We also collected photometry in four MIRI broadband filters: F1000W, F1500W, F1800W, and F2100W ($\lambda_\mathrm{pivot}$= 9.954, 15.065, 17.987, 20.795 \mic, respectively). For the observations we used the FASTR1 readout and a 4-point dither, with 5 as our selected starting dither position.

Exposure times for each object were chosen via the Exposure Time Calculator using a model spectrum \citep{marley_sonora_2021} scaled to the object's [4.5] magnitude. We required that the MIRI photometry achieve a signal-to-noise ratio of 30 and the NIRSpec and MIRI LRS simulated observations achieve a median signal-to-noise ratio of 30, allowing us to achieve $\sigma_{F_\mathrm{bol}}/F_\mathrm{bol}\leq0.03$. We also required a signal-to-noise ratio of 10 at both 5 and 12 \mic, as these points are where we merge our spectra and photometry when we build the spectral energy distributions. These signal-to-noise ratio requirements have all been surpassed due to JWST surpassing pre-launch requirements. Details on individual observations can be found in Table \ref{tbl:obs}.

\begin{deluxetable*}{lcccccccccccccc}
\tablecaption{Target Exposure Times\label{tbl:obs}}
\tablehead{
\colhead{Name} &
\multicolumn{2}{c}{NIRSpec} &
\multicolumn{2}{c}{MIRI LRS} &
\multicolumn{2}{c}{MIRI F1000W} &
\multicolumn{2}{c}{MIRI F1500W} &
\multicolumn{2}{c}{MIRI F1800W} &
\multicolumn{2}{c}{MIRI F2100W} &
\\
\cmidrule(lr){2-3} \cmidrule(lr){4-5} \cmidrule(lr){6-7} \cmidrule(lr){8-9} \cmidrule(lr){10-11} \cmidrule(lr){12-13}
\colhead{} &
\colhead{Obs. Date} &
\colhead{N$\times$Int} &
\colhead{Obs. Date} &
\colhead{N$\times$Int} &
\colhead{Obs. Date} &
\colhead{N$\times$Int} &
\colhead{Obs. Date} &
\colhead{N$\times$Int} &
\colhead{Obs. Date} &
\colhead{N$\times$Int} &
\colhead{Obs. Date} &
\colhead{N$\times$Int} &
\\
\colhead{} &
\colhead{(UT)} &
\colhead{(s)} &
\colhead{(UT)} &
\colhead{(s)} &
\colhead{(UT)} &
\colhead{(s)} &
\colhead{(UT)} &
\colhead{(s)} &
\colhead{(UT)} &
\colhead{(s)} &
\colhead{(UT)} &
\colhead{(s)} &
}
\rotate
\startdata
WISE J0247+37& 2022-09-21&5$\times$  10.93&2022-09-21&2$\times$ 249.75&2022-09-21&4$\times$  13.88&2022-09-21&4$\times$  13.88&2022-09-21&4$\times$  33.30&2022-09-21&8$\times$  76.31\\
WISEPA J0313+78&2022-09-17&5$\times$   9.37&2022-09-17&2$\times$  77.70&2022-09-17&4$\times$  13.88&2022-09-17&4$\times$  13.88&2022-09-17&4$\times$  13.88&2022-09-17&4$\times$  19.43\\
WISE J0359$-$54& 2022-09-12&5$\times$  17.16&2022-09-12&2$\times$1387.52&2022-09-12&4$\times$  19.43&2022-09-12&4$\times$  19.43&2022-09-12&4$\times$  49.95&2022-09-12&12$\times$  79.55\\
WISE J0430+46& 2022-09-21&5$\times$   9.37&2022-09-21&2$\times$ 180.38&2022-09-21&4$\times$  13.88&2022-09-21&4$\times$  13.88&2022-09-21&4$\times$  19.43&2022-09-21&4$\times$  52.73\\
WISE J0535$-$75& 2022-09-19&5$\times$  15.60&2022-09-13&2$\times$ 693.76&2022-09-19&4$\times$  16.65&2022-09-19&4$\times$  19.43&2022-09-19&4$\times$  44.40&2022-09-19&12$\times$  74.00\\
WISE J0734$-$71& 2023-07-09&5$\times$  17.16&2023-07-09&2$\times$1248.77&2023-07-09&4$\times$  16.65&2023-07-09&4$\times$  19.43&2023-07-09&4$\times$  47.18&2023-07-09&12$\times$  76.78\\
WISE J0825+28& 2022-10-25&5$\times$  12.48&2022-10-25&2$\times$ 416.26&2022-10-29&4$\times$  13.88&2022-10-29&4$\times$  13.88&2022-10-29&4$\times$  19.43&2022-10-29&4$\times$  44.40\\
ULAS J1029+09& 2023-05-04&5$\times$  10.93&2023-05-04&2$\times$ 208.13&2023-05-04&4$\times$  13.88&2023-05-04&4$\times$  13.88&2023-05-04&4$\times$  27.75&2023-05-04&8$\times$  65.21\\
CWISEP J1047+54&2022-11-25&5$\times$  34.30&2022-11-11&10$\times$1681.12&2022-11-11&4$\times$  41.63&2022-11-11&4$\times$  44.40&2022-11-11&12$\times$  62.90&2022-11-11&32$\times$ 105.10\\
WISE J1206+84& 2022-10-03&5$\times$  17.16&2022-10-17&2$\times$ 971.26&2023-02-27&4$\times$  16.65&2022-10-29&4$\times$  19.43&2022-10-29&4$\times$  41.63& 2023-02-27&8$\times$  98.51\\
SDSSP J1346$-$00 &2023-07-18&5$\times$   9.37&2023-07-18&2$\times$  61.05&2023-07-17&4$\times$  13.88&2023-07-17&4$\times$  13.88&2023-07-17&4$\times$  19.43&2023-07-17&4$\times$  49.95\\
WISEPC1405+55 &2023-02-28&5$\times$   9.37&2023-03-04&2$\times$ 194.25&2023-05-03&4$\times$  13.88&2023-05-03&4$\times$  13.88&2023-05-03&4$\times$  13.88&2023-05-03&4$\times$  24.98\\
CWISEP J1446$-$23&2023-02-03&5$\times$  24.95&2023-02-25&6$\times$1181.24&2023-02-25&4$\times$  30.52&2023-02-25&4$\times$  27.75&2023-02-25&4$\times$  80.48&2023-02-25&16$\times$  79.78\\
WISE J1501$-$40& 2023-02-25&5$\times$   9.37&2023-03-07&2$\times$ 111.00&2023-03-05&4$\times$  13.88&2023-03-05&4$\times$  13.88&2023-03-05&4$\times$  22.20&2023-03-05&4$\times$  83.25\\
WISEPA J1541$-$22&2023-03-05&5$\times$   9.37&2023-03-05&2$\times$ 208.13&2023-03-05&4$\times$  13.88&2023-03-05&4$\times$  13.88&2023-03-05&4$\times$  13.88&2023-03-05&8$\times$  26.36\\
SDSS J1624+00& 2023-02-25&5$\times$   9.37&2023-03-08&2$\times$  41.63&2023-03-11&4$\times$  13.88&2023-03-11&4$\times$  13.88&2023-03-11&4$\times$  13.88&2023-03-11&4$\times$  24.98\\
WISEPA J1959$-$33&2022-09-23&5$\times$   9.37&2022-09-28&2$\times$  83.25&2022-09-28&4$\times$  13.88&2022-09-28&4$\times$  13.88&2022-09-28&4$\times$  16.65&2022-09-28&4$\times$  47.18\\
WISEPC J2056+14&2022-09-20&5$\times$   9.37&2022-09-20&2$\times$ 138.75&2022-09-20&4$\times$  13.88&2022-09-20&4$\times$  13.88&2022-09-20&4$\times$  13.88&2022-09-20&4$\times$  24.98\\
WISE J2102$-$44& 2022-09-19&5$\times$   9.37&2022-09-19&2$\times$ 180.38&2022-09-19&4$\times$  13.88&2022-09-19&4$\times$  13.88&2022-09-19&4$\times$  19.43&2022-09-19&4$\times$  61.05\\
WISE J2209+27& 2022-10-17&5$\times$  10.93&2022-10-17&2$\times$ 513.38&2022-10-17&4$\times$  13.88&2022-10-17&4$\times$  13.88&2022-10-17&4$\times$  19.43&2022-10-17&4$\times$  41.63\\
WISEA J2354+02 &2023-07-06&5$\times$  14.04&2023-07-06&2$\times$ 582.76&2023-07-06&4$\times$  16.65&2023-07-06&4$\times$  13.88&2023-07-06&4$\times$  24.98&2023-07-06&4$\times$  83.25\\
WISEA J2018$-$74 & -- & -- &2022-09-19&2$\times$  61.05&2022-09-19&4$\times$  13.88&2022-09-19&4$\times$  13.88&2022-09-19&4$\times$  16.65&2022-09-19&4$\times$  38.85\\
WISEPA J0410+15&--&--&--&--&2022-09-22&4$\times$  13.88&2022-09-22&4$\times$  13.88&2022-09-22&4$\times$  13.88&2022-09-22&4$\times$  27.75\\
\enddata
\end{deluxetable*}

Two of our objects had issues with their target acquisition which resulted in a portion of their observations failing, which is why they are colored red in Figure \ref{fig:Sample}. WISEPA J0410+15 (Y0) is missing both spectral observations due to an error in the submitted proper motion that caused it to fall outside of the acquisition area. Its photometry is reported in this paper, but this object is not included in our analysis. WISEPA J2018$-$74 (T7) is missing its NIRSpec observation due to an failed guide star acquisition. The reason for its failure is unknown and the failure was not identified until after the period to request repeating the observation had passed. We use a previously observed near-infrared spectrum and Spitzer photometry as a replacement for the NIRSpec data, and we calculate a correction for the lack of NIRSpec data in Section \ref{sec:Varients}. Guide star failures did occur for several of our other targets, again with no clear cause, but the others were identified quickly and the observations were repeated upon request.

\section{Data Reduction} \label{sec:reduc}
\subsection{JWST Pipeline}
We used the JWST Science Calibration Pipeline (Version 1.12.5) for the data reduction, using the 11.17.6 Calibration Reference Data System (CRDS) version and the 1193.pmap CRDS context to assign the reference files. There were three issues we noted while processing the data:

1. CWISEP J1047+54 is not identified by the ``source\_catalog" step in the F2100W filter due to low signal, even though it could be seen by eye. This was overcome by lowering the signal-to-noise threshold for detection (``snr\_threshold'') of this step from 3.0 to 1.0.

2. The uncertainty due to the photometric zero point of the MIRI photometry is not included in the pipeline. The current version of the pipeline for MIRI photometry is accurate to 3\% or better based on measured flux calibration stars\footnote{https://www.stsci.edu/contents/news/jwst/2023/updates-to-the-miri-imager-flux-calibration-reference-files?page=3\&filterUUID=0655d914-43ee-4d09-a91b-9f45be575098}, so we adopt a 3\% uncertainty for our MIRI photometric points.

3. There is a slight error in the wavelength calibration past 12 \mic, the nominal end of the MIRI LRS spectrum, but the error is small enough (0.01$-$0.03 \mic) and over a short enough range that we do not attempt correction. This wavelength error is the residual from the wavelength error that previously plagued the entire MIRI LRS spectrum \citep[e.g.][]{beiler_first_2023} before a correction was added to the JWST pipeline after further calibration observations. Further calibration will likely be required to improve the wavelength calibration past 12 \mic, but likely is not a high priority since this range is outside of the nominal range of the MIRI LRS spectrum.

\subsection{Pipeline Calibration} \label{subsec:PipeCali}
The pre-flight goal for the precision of the absolute flux calibration of JWST spectra was $\sim$10\% \citep{gordon_james_2022}. We use our sample to get a sense for the reliability of the pipeline's baseline absolute flux calibration by comparing synthetic photometry to previous observations. 
\subsubsection{NIRSpec}
\begin{figure}
\includegraphics[width=.45\textwidth]{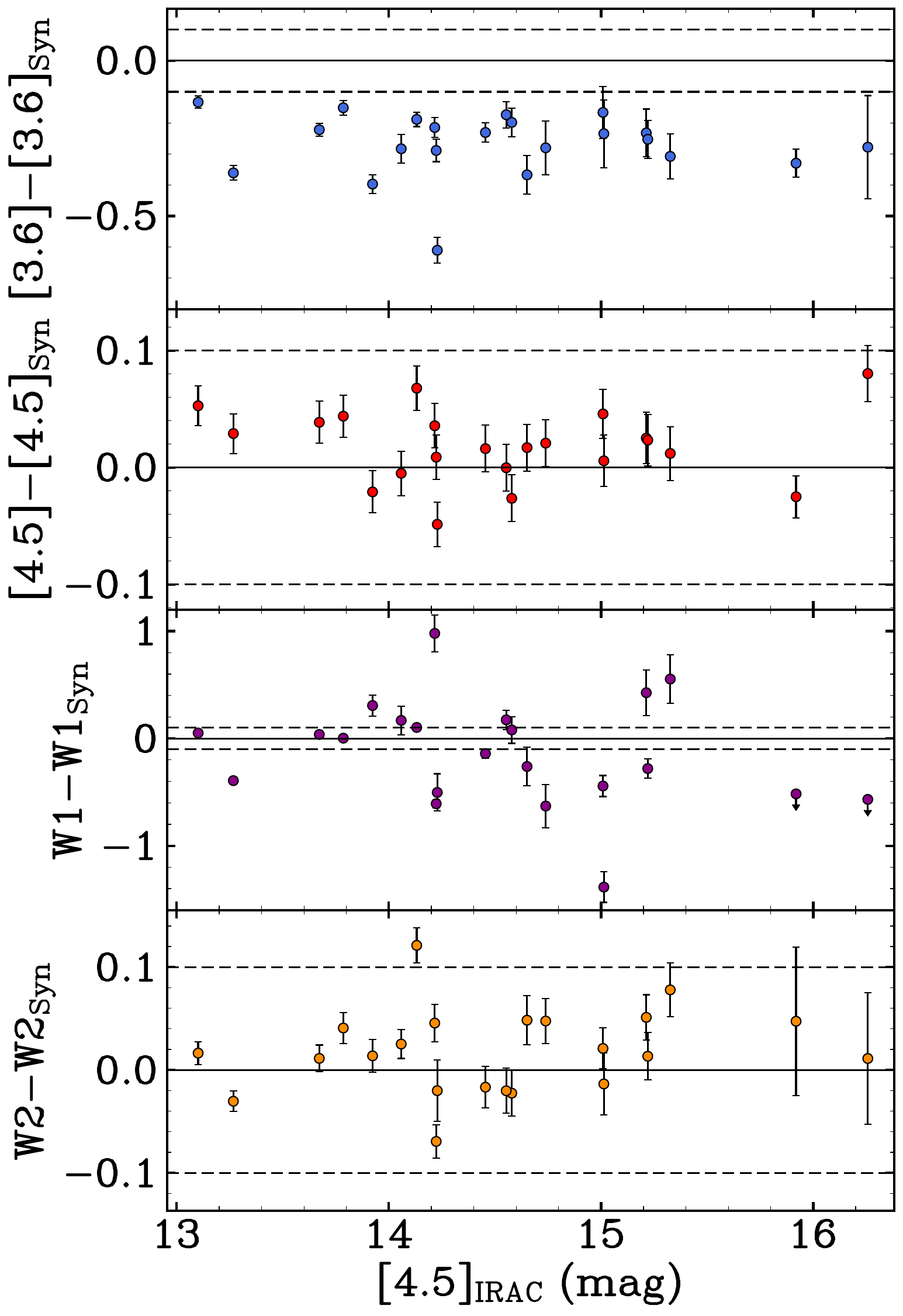}
\centering
\caption{The difference between observed and synthetic magnitudes (Spitzer and WISE) for our sample. Note the y-axis ranges are all different. Therefore we added a solid line and dashed lines at 0 and $\pm$ 0.1 magnitudes respectively for ease of comparison between the different photometric bands while retaining the fine detail for each distribution. The [4.5], W1, and W2 synthetic photometries are consistent with observations with varying levels of scatter, but the [3.6] synthetic photometry is systematically $\sim$0.3 magnitudes fainter than the observed magnitude. Since this is not the case for the W1 photometry (which covers a similar wavelength range) or the [4.5] Spitzer photometry, this is likely an issue with the observed [3.6] magnitudes, possibly due to a light leak.} \label{fig:PhotoUncal}
\end{figure}

We generate synthetic WISE (W1 and W2) and Spitzer ([3.6] and [4.5]) photometry from the NIRSpec spectra following methods and using zero-points from \citet{reach_absolute_2005}, \citet{cushing_spitzer_2006}, and \citet{jarrett_spitzer-wise_2011}. Figure \ref{fig:PhotoUncal} shows the difference between the observed (found in Table {\ref{tbl:prop}) and synthetic photometry, as a function of [4.5] apparent magnitude. Variability is not a concern for these objects, as \citet{brooks_long-term_2023} detected no long-term variability above 20 mmags for late-T and Y dwarfs at these wavelengths, and \citet{trucks_variability_2019} observed a maximum short-term variability of 4.2\%. With this large sample we would also expect any variability to impact the scatter of the data, but not the average offset.} The deltas of the W2 and [4.5] filters, which cover a similar wavelength range, are both evenly distributed around 0, with a scatter of approximately twice the individual uncertainties (0.1 and 0.05 magnitudes of scatter respectively). The W1 deltas are also evenly distributed around 0, but with a much larger scatter. These three trends point to the pipeline calibration being well within the 10\% pre-flight goal.

In contrast to the W1, W2, and [4.5] filters, the [3.6] filter stands out with the synthetic photometric magnitudes being consistently $\sim$0.3 magnitudes fainter than observed magnitudes. This has previously been seen in observations of WISE J0359$-$54 \citep{beiler_first_2023} and WISE J085510.83$-$071442.5 \citep{luhman_jwstnirspec_2023}, with WISE J0855$-$07 showing a $\sim$0.4 mag offset for the synthetic photometry of both NIRSpec PRISM and G395M. The fault seems to be with the observed [3.6] magnitudes since this offset is occurring for two different NIRSpec modes, is not occurring for W1 even though it covers a similar wavelength regime, and is the only one of the filters we test with this issue. 

\begin{figure}
\includegraphics[width=.45\textwidth]{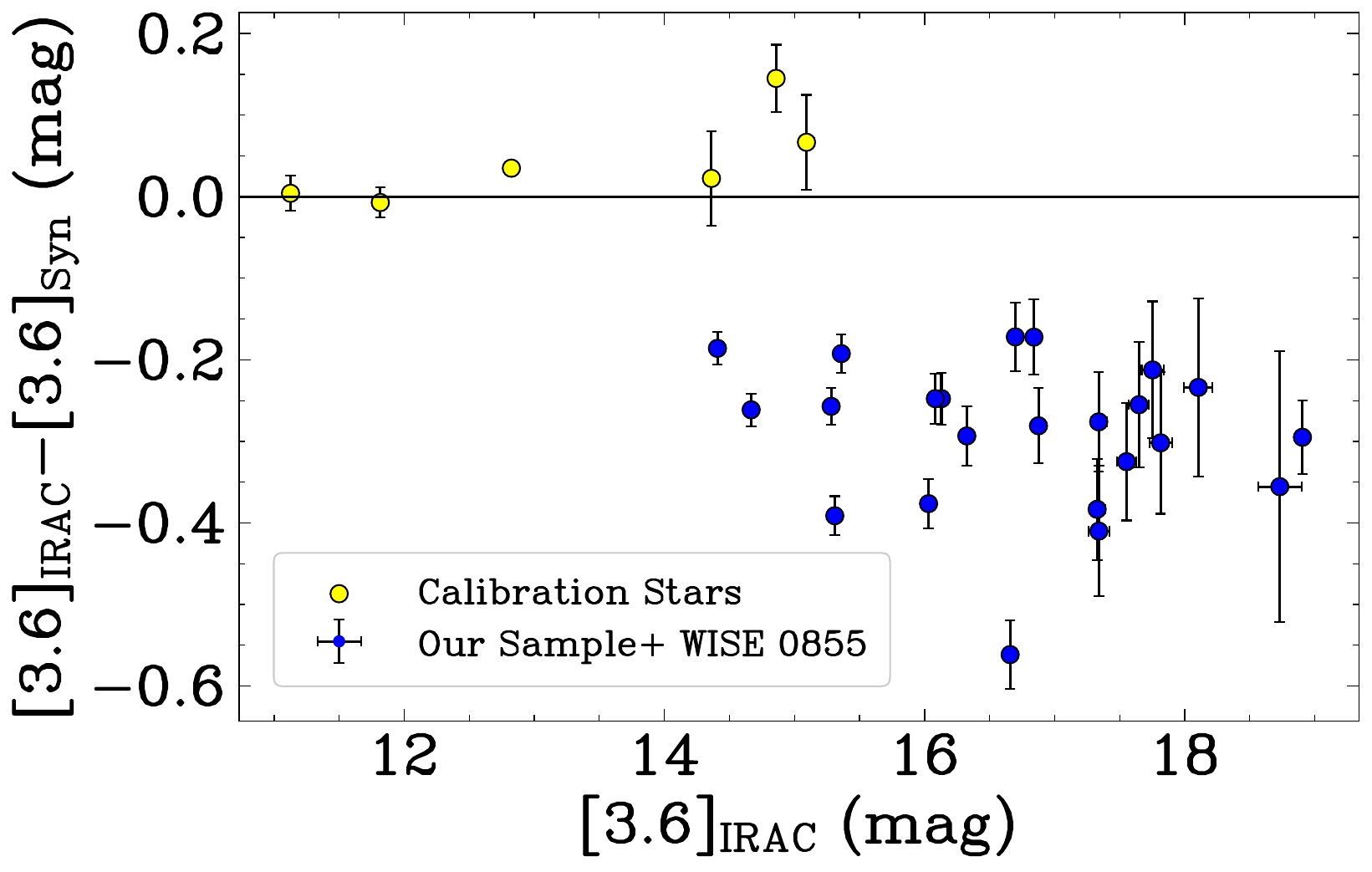}
\centering
\caption{The difference between the observed and synthetic [3.6] magnitudes for both our sample (blue) and 6 JWST calibrator stars (yellow). The calibrator stars are near zero, distinct from the $\sim$0.3 magnitude systematic offset of our sample. This means the [3.6] offset is likely due to a trait inherent to ultracool objects, possibly their extreme red spectra in this bandpass.} \label{fig:Ch1Offset}
\end{figure}

To see if this issue extends beyond substellar observations, we found 6 objects selected as calibrators for JWST which have both available NIRSpec PRISM/CLEAR spectra and [3.6] magnitudes \citep{krick_spitzer_2021}. Their observations and properties are listed in Table \ref{tbl:cal}. These NIRSpec observations were taken using a different subarray with a shorter height (32 vs 64 pixels), which unfortunately means that there is some overlap between nods when performing background subtraction. This results in a poor out-of-the-box flux calibration, and as such we must first calibrate these 6 spectra to their K-band magnitudes \citep{cutri_vizier_2003,lawrence_ukirt_2007, mcmahon_first_2013, dye_ukirt_2018}, which we also list in Table \ref{tbl:cal}. We then calculate their synthetic [3.6] magnitudes and the deltas to observed values, and plot them along with the data for our sample and WISE 0855 ([3.6] = 17.34, [3.6]$_\mathrm{IRAC}-[3.6]_\mathrm{Sim}= -0.41$, \citep{luhman_jwstnirspec_2023}) in Figure \ref{fig:Ch1Offset}. We plot these deltas as a function of [3.6] magnitude since some of the calibrator stars lack [4.5] measurements. The uncertainties on the deltas for the calibrator stars include the K-band magnitude uncertainties.

\begin{deluxetable*}{lccccccc}
\tablecaption{Calibrator Star Observations and Properties\label{tbl:cal}}
\tablehead{
\colhead{Obj. Name} &
\colhead{Spec. Type\tablenotemark{a}} &
\colhead{Program \#} &
\colhead{K-band (mag)} &
\colhead{IRAC [3.6]\tablenotemark{a} (mag)} &
\colhead{Mode} &
\colhead{Slit} &
\colhead{Subarray} 
}
\startdata
2MASS J18022716+6043356   & A2V& 1536 & 11.83$\pm$0.018\tablenotemark{b}& 11.828$\pm$0.008 & PRISM/CLEAR & S1600A1 & SUB512 \\
2MASS J17571324+6703409   & A3V& 4496 & 11.16$\pm$0.023\tablenotemark{b}& 11.150$\pm$0.005 & PRISM/CLEAR & S1600A1 & SUB512 \\
G 191-B2B & DA1                & 1537 & 12.78$\pm$0.003\tablenotemark{c}& 12.835$\pm$0.007 & PRISM/CLEAR & S1600A1 & SUB512 \\
GD 153 & DA1                   & 4497 & 14.31$\pm$0.062\tablenotemark{b}& 14.370$\pm$0.007 & PRISM/CLEAR & S1600A1 & SUB512 \\
LDS 749B & DBQ4                & 4497 & 15.11$\pm$0.012\tablenotemark{d}& 15.092$\pm$0.057 & PRISM/CLEAR & S1600A1 & SUB512 \\
2MASS J03323287$-$2751483 & F7V& 4498 & 15.06$\pm$0.012\tablenotemark{e}& 14.859$\pm$0.040 & PRISM/CLEAR& S1600A1 & SUB512\\
\enddata
\tablenotetext{a}{All spectral types and IRAC [3.6] magnitudes come from \citep{krick_spitzer_2021}.}
\tablenotetext{b}{2MASS K$_{s}$ magnitude from the Two Micron All Sky Survey \citep{cutri_vizier_2003}.}
\tablenotetext{c}{UKIRT WFCAM K (MKO) magnitude from the UKIRT Hemisphere Survey \citep{dye_ukirt_2018}.}
\tablenotetext{d}{VISTA K$_{s}$ (2MASS) magnitude from the VISTA Hemisphere Survey \citep{lawrence_ukirt_2007}.}
\tablenotetext{e}{UKIRT WFCAM K (MKO) magnitude from the UKIRT Infrared Deep Sky Survey \citep{mcmahon_first_2013}.}
\end{deluxetable*}

The differences for all the calibrator stars are near zero, which is different from the systematic offset we are seeing in the late brown dwarfs. This means the offset we are seeing is likely a result of the strange shape of brown dwarf spectra within the [3.6] bandpass, which starts with near zero flux at the blue edge and climbs steeply towards the red edge. As such, a light leak or a small discrepancy in the red edge of the [3.6] transmission curve could explain the trend we are seeing. Within this current range of spectral types, the offset is not obviously a function of spectral type.  Future NIRSpec observations of M, L, and early T dwarfs would be useful in order to see how this offset changes as a function of spectral type and morphology. 

\subsubsection{MIRI LRS}
\begin{figure}
\includegraphics[width=.45\textwidth]{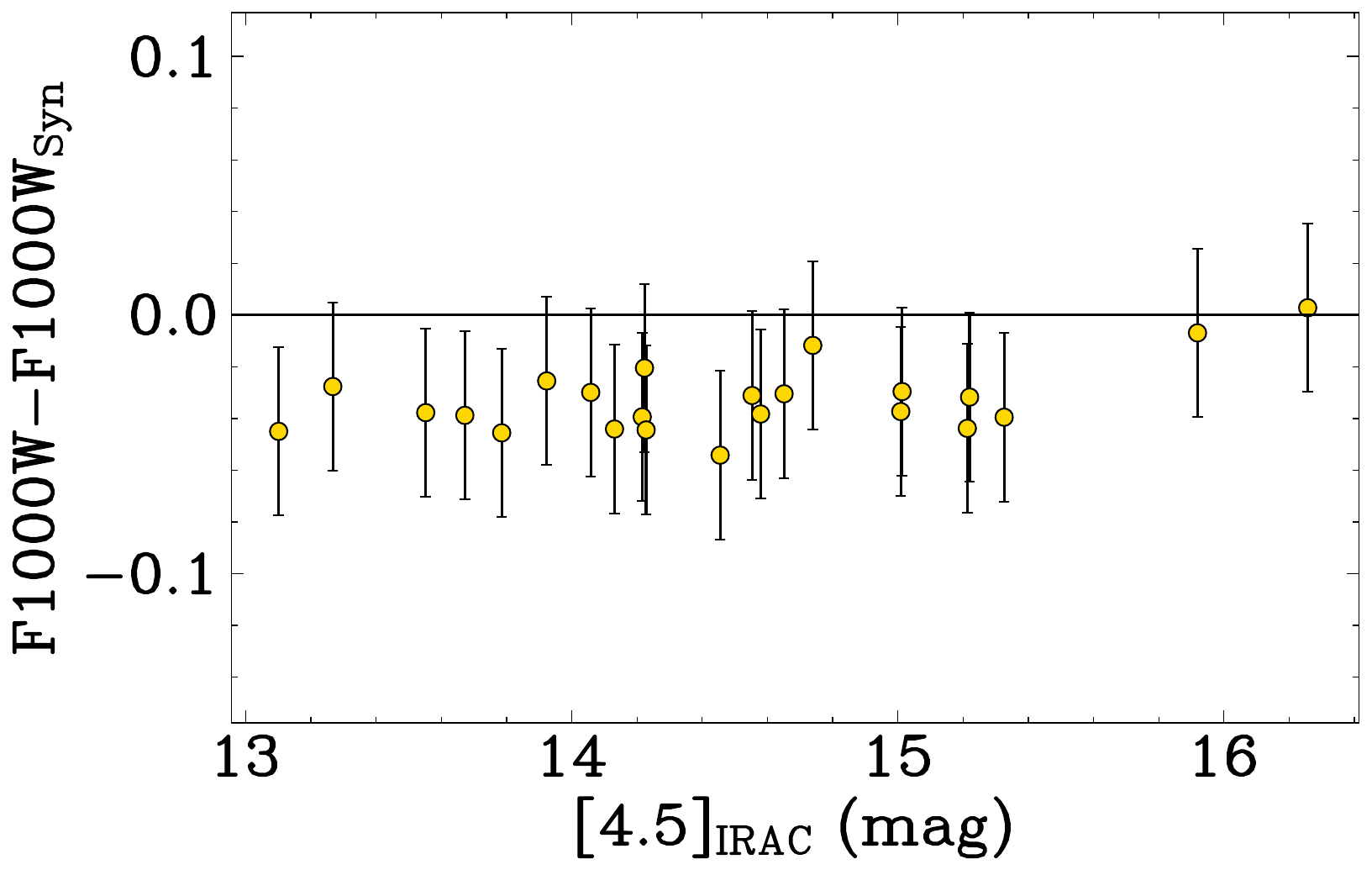}
\centering
\caption{The difference of between the observed and synthetic F1000W magnitudes for our sample, which show synthetic photometry is $\sim$0.04 magnitudes dimmer. This is likely due to the JWST spectral pipeline pathloss correction step not accounting for the actual slit position.} \label{fig:PhotoUncalF1000W}
\end{figure}

We also test the pipeline's flux calibration of our MIRI LRS data using synthetic F1000W photometry, which we generate following the procedure in \citet{gordon_james_2022}. These are plotted in Figure \ref{fig:PhotoUncalF1000W}. We see a systematic offset, with the simulated photometry being $\sim$0.04 magnitudes fainter than the observed F1000W magnitude, just over the uncertainty of our photometric observations. Unfortunately, F1000W is the only observed photometry whose entire bandpass fits within the MIRI LRS spectral range, so we cannot determine whether this is an issue with the photometric or spectroscopic absolute flux calibration. The JWST spectral pipeline currently does not include true pathloss correction based on the actual slit point, so this is likely the culprit, and a 4\% error is within expectations for the current absolute flux calibration of MIRI LRS. 

\subsection{Flux Calibration and Merging of Spectra} 
\begin{figure*}
\includegraphics[width=\textwidth]{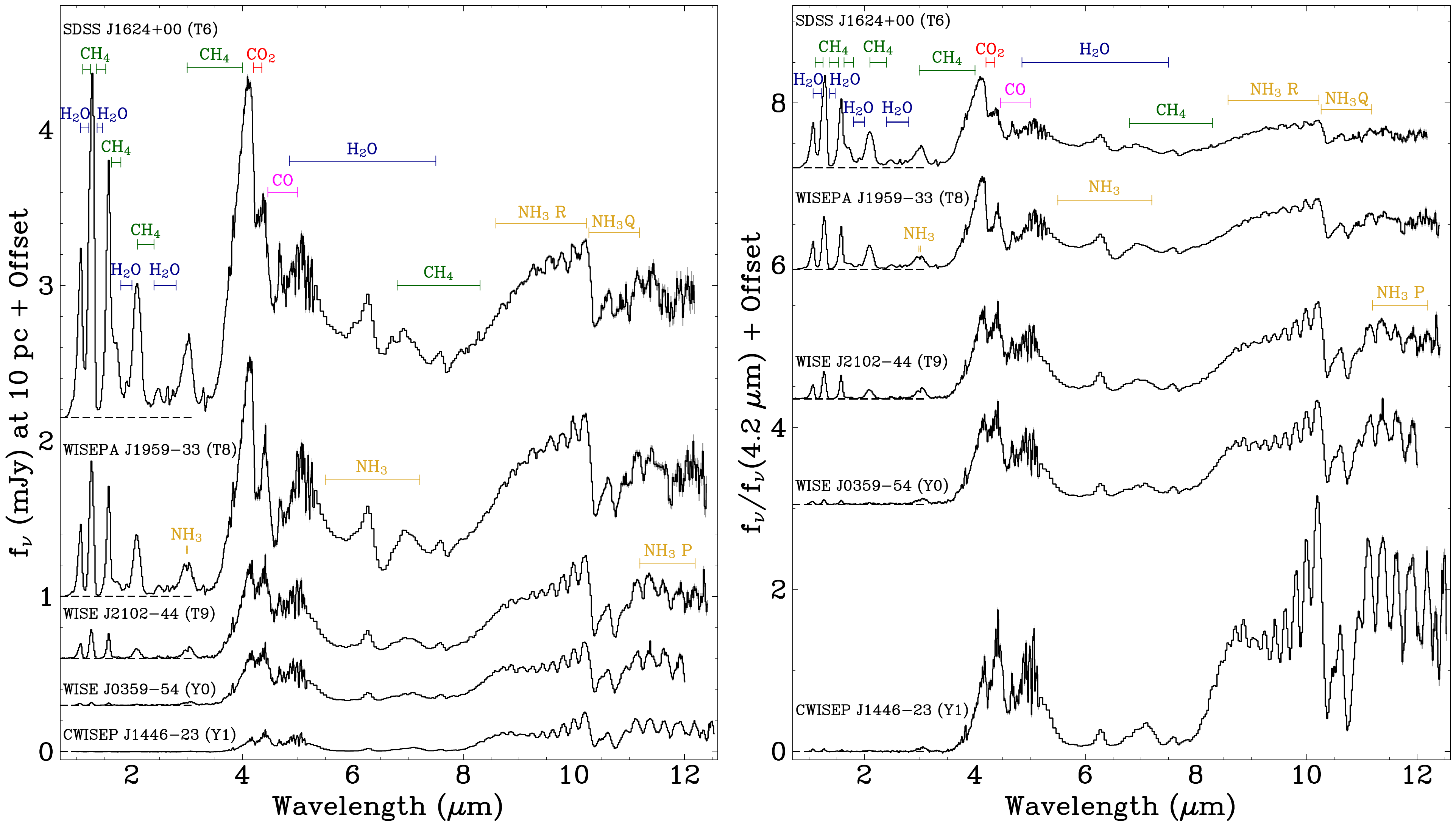}
\centering
\caption{A spectral sequence of representative objects, showing how varied the spectra are for these late types, plotted both in absolute flux (left) and normalized flux (right), and offset for clarity. The zero flux level for each spectrum is marked by the dashed line at the left edge of each spectrum.} \label{fig:AbsorptionSequenceAbsNorm}
\end{figure*}

While the absolute flux calibration of JWST appears to be within the pre-flight goal of $\sim$10\%, higher precision calibration is needed to achieve our desired relative precision of 3\% on the bolometric flux measurement. We therefore obtained photometry to improve the absolute calibration to at least $\sim$5\%. The NIRSpec spectra were absolutely calibrated with [4.5] photometry \citep{kirkpatrick_field_2021}, and the MIRI LRS spectra were calibrated with MIRI F1000W photometry. Both of these photometric values and uncertainties can be found in Table {\ref{tbl:prop}. From these magnitudes we calculate the scaling factors needed to convert our spectra and their uncertainties to absolute units of Jansky \citep{reach_absolute_2005,gordon_james_2022}. It is worth noting Spitzer and JWST photometry assume different nominal spectra, $\nu f_{\nu}$ = constant and $f_\lambda$ = constant, respectively.  

We create continuous $\sim$0.6--$\sim$13 \mic{} spectra by merging the NIRSpec and MIRI spectra at their overlap from 5 to 5.3 \mic. The NIRSpec spectra have a higher resolving power than the MIRI LRS spectra at these wavelengths but a lower signal-to-noise ratio. To balance the trade-off between including spectral features only visible in NIRSpec and including low signal-to-noise data, we cut the NIRSpec spectra where they first drop below a signal-to-noise ratio of 10; the precise wavelength at which this occurs varies across our sample. We also trim the spectra longward of 12 \mic{} at the shortest wavelength point where the signal-to-noise ratio first drops below 10, and similarly trim at the longest wavelength point with negative flux before the $Y$ band ($<$0.96 \mic). As such, the upper and lower wavelength bounds of the merged spectra vary for each object. The final spectral energy distributions for each object, including MIRI photometry, are plotted in Figures \ref{fig:AllSpecLog1}$-$\ref{fig:AllSpecLog3}. With these flux calibrated spectra, we are able to calculate synthetic photometry for most of the JWST photometric filters. We report this process and its results in Appendix \ref{sec:app}.

\begin{figure*}
\includegraphics[angle=90,width=\textwidth]{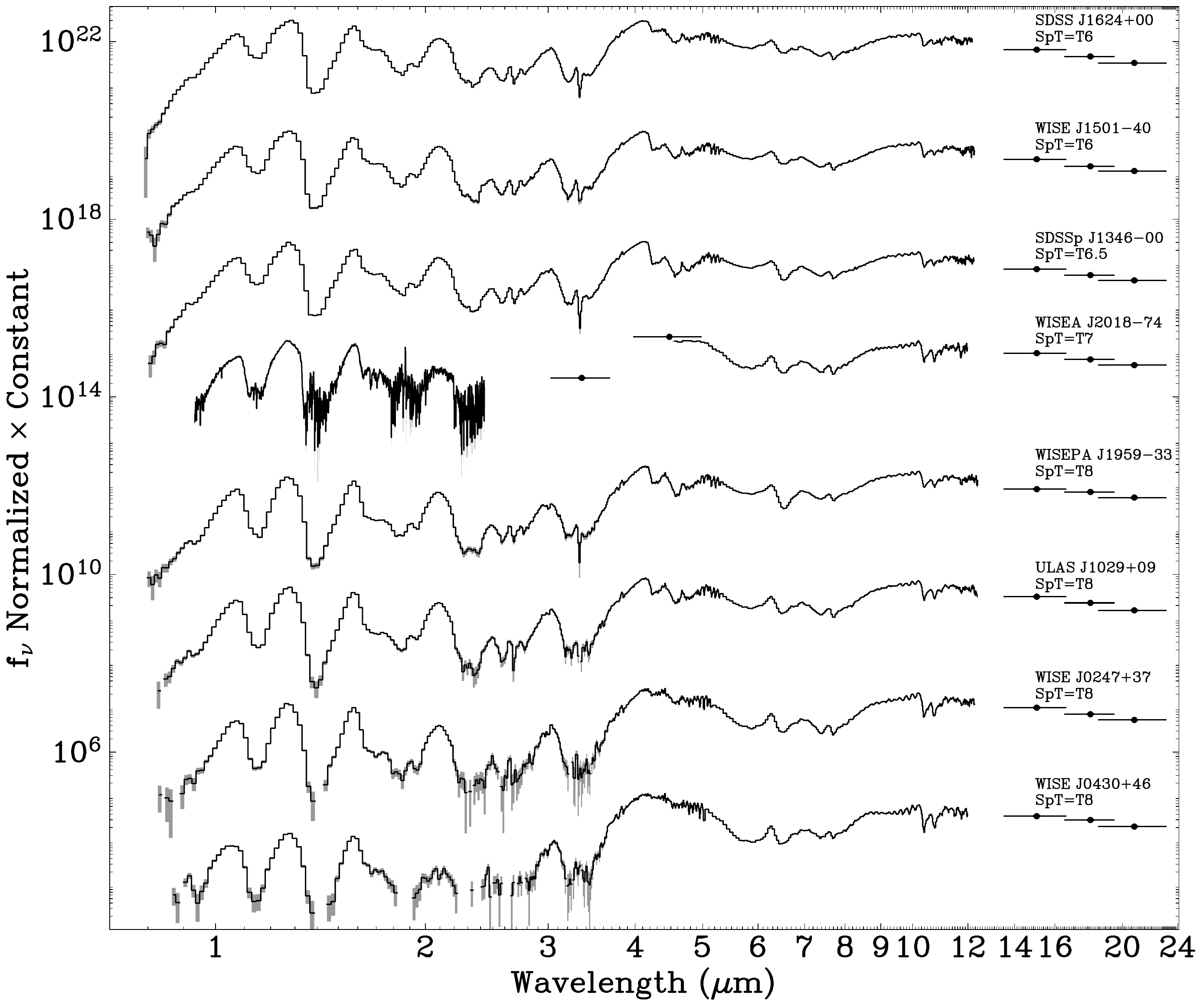}
\centering
\caption{The reduced and calibrated spectral energy distributions of our sample plotted in normalized f$_\nu$, primarily ordered by spectral type and secondarily ordered by [4.5] absolute magnitude. We removed points which are consistent with 0 (f$_i<\sigma_i$), and show the uncertainties of our spectrum with the grey shaded bars. Note that WISEA J2018$-$74 lacks NIRSpec data, and so we instead plot the near-infrared Magellan/FIRE spectrum from \citep{burgasser_fire_2011} and Spitzer [3.6] and [4.5] photometry.}\label{fig:AllSpecLog1}
\end{figure*}

\begin{figure*}
\includegraphics[angle=90,width=\textwidth]{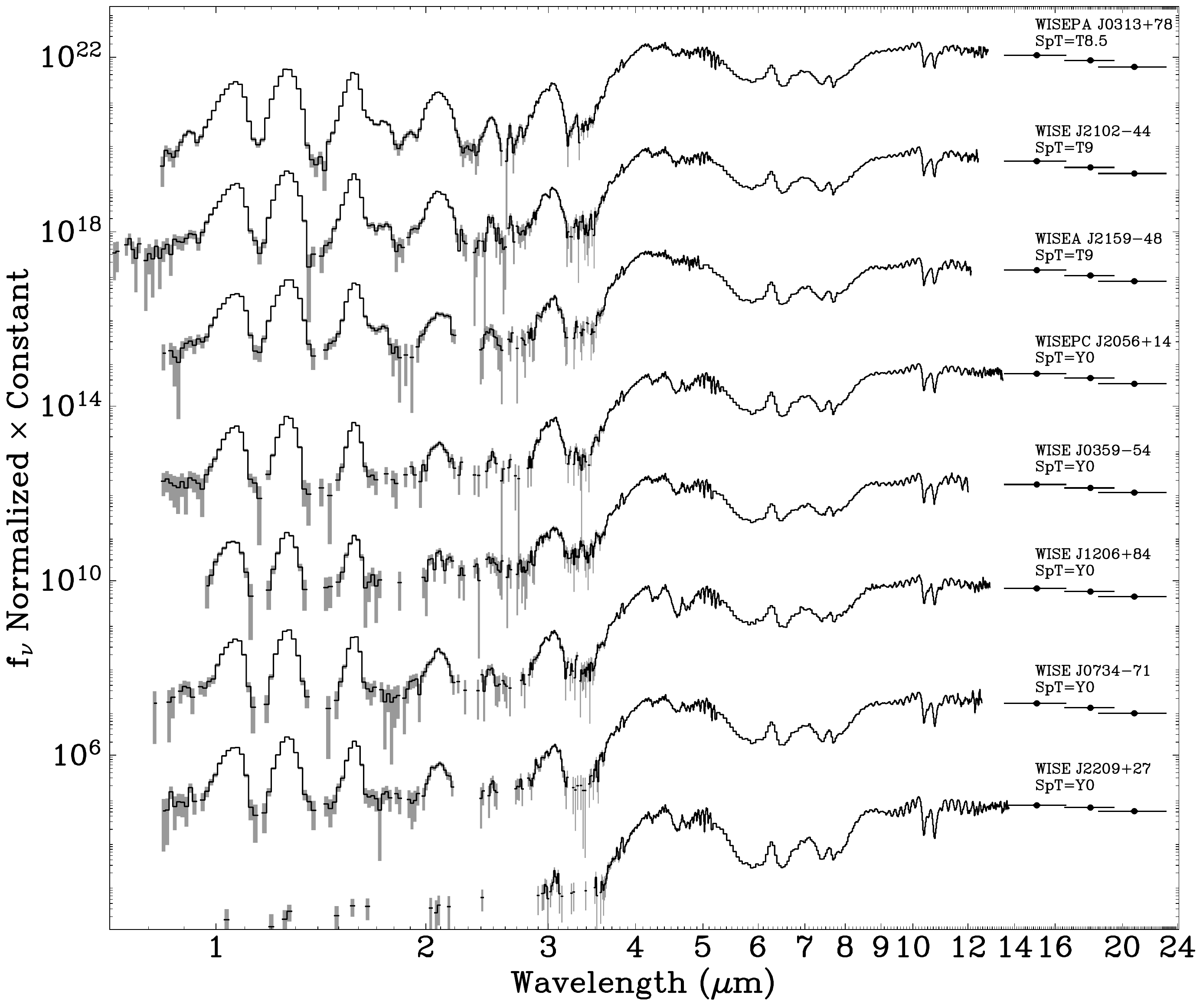}
\centering
\caption{Same as Figure \ref{fig:AllSpecLog1}.} \label{fig:AllSpecLog2}
\end{figure*}

\begin{figure*}
\includegraphics[angle=90,width=\textwidth]{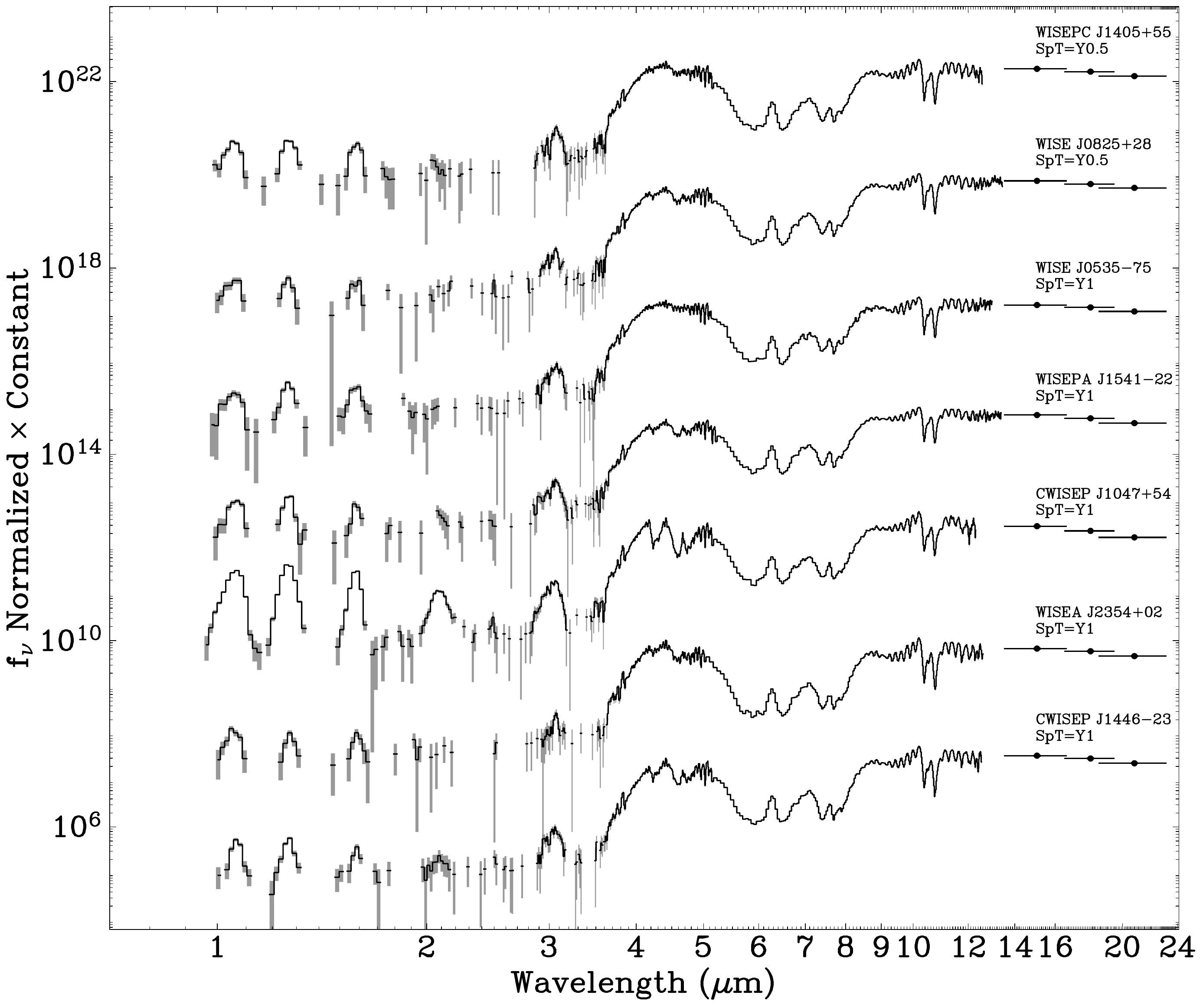}
\centering
\caption{Same as Figure \ref{fig:AllSpecLog1}.} \label{fig:AllSpecLog3}
\end{figure*}

\section{Near-Infrared Spectral Typing}\label{sec:nirSpecType}
Two of our objects, CWISEP J1047+54 and CWISEP J1446$-$23, have no previous near-infrared spectroscopic observations, and as such only have spectral type estimates inferred from their Spitzer [3.6]$-$[4.5] colors \citep[Y0 and $\geq$Y1 respectively,][]{meisner_expanding_2020}. We determined the spectral types for these two objects by comparing their near-infrared spectra to spectral standards defined in \citet{cushing_discovery_2011} and \citet{kirkpatrick_further_2012}. Since the spectral standards have a higher resolution than our near-infrared data, we convolve the standards down to the resolution of NIRSpec for ease of comparison, which can be seen in Figures \ref{fig:SpecType}.

\begin{figure*}
\includegraphics[width=1\textwidth]{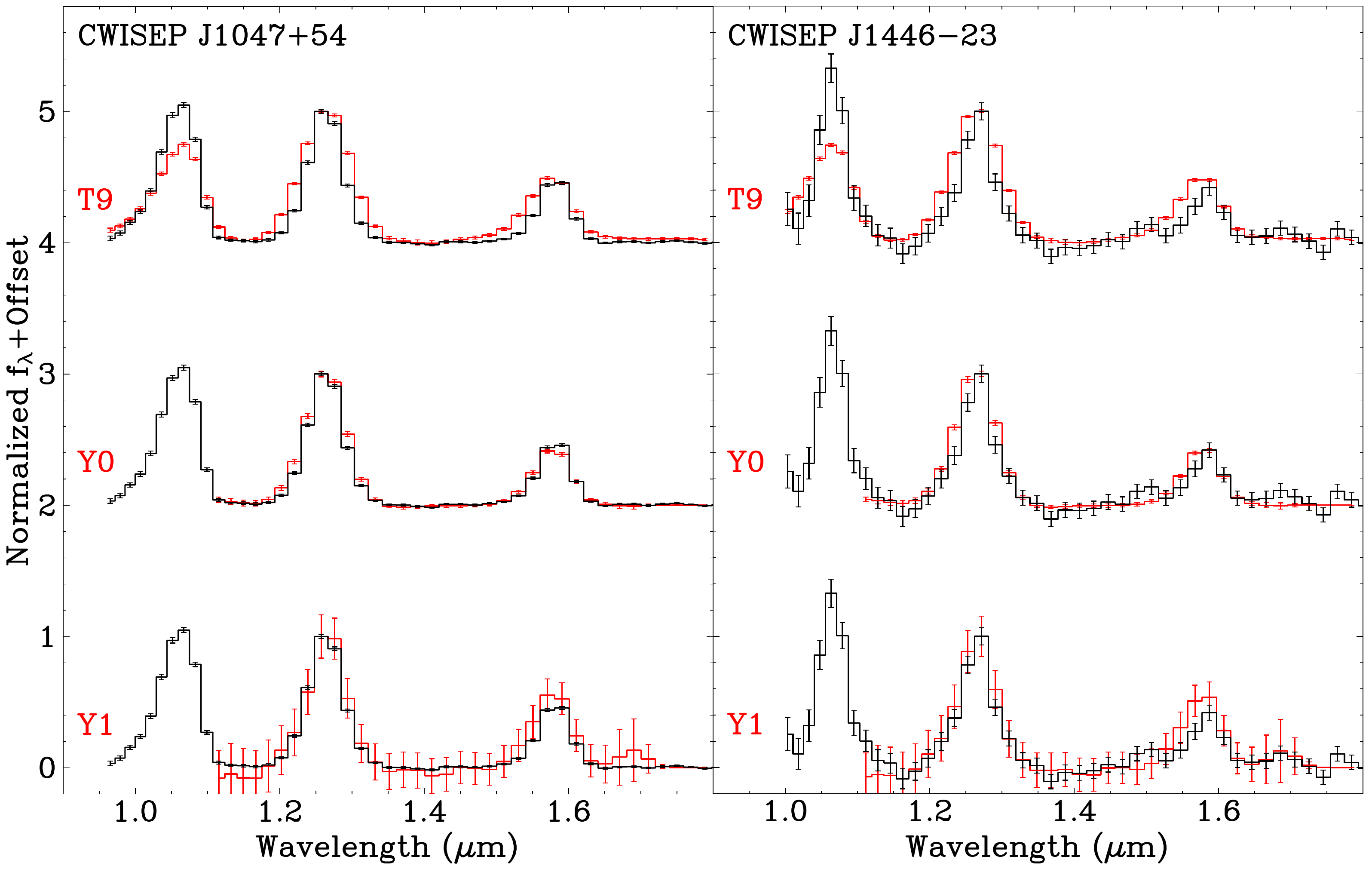}
\centering
\caption{Near-infrared portion of the NIRSpec spectra of CWISEP J1047+54 (left, black) and CWISEP J1446$-$23 (right, black) compared to the spectral standards (red), all of which are normalized to unity at the peak of the $J$ band and offset by constants. The spectra for the spectral standards come from \citet{lucas_discovery_2010} (T dwarf) and \citet{schneider_hubble_2015} (Y dwarfs). We classify both of our objects as a Y1 dwarfs.} \label{fig:SpecType}
\end{figure*}

Both objects are clearly Y dwarfs, as they exhibit strong $Y$-band peaks compared to the T dwarf spectral standards. We classify both CWISEP J1047+54 and CWISEP J1446$-$23 as Y1 dwarfs primarily due to the narrowness of their $J$-band peaks, however, it is difficult to precisely spectral type these objects due to the low signal-to-noise ratio of the Y1 spectral standard and the resolution of our data. 

\section{Prominent Absorption Bands}\label{sec:spec}
With such a broad spectral range we are able to see a plethora of molecular absorption features, including water ($\mathrm{H}_2\mathrm{O}$), methane ($\mathrm{CH}_4$), ammonia ($\mathrm{NH}_3$), carbon monoxide ($\mathrm{CO}$), and carbon dioxide ($\mathrm{CO}_2$). These features have been seen in previously published Spitzer, JWST, and ground-based observations \citep{oppenheimer_spectrum_1998,roellig_spitzer_2004,schneider_hubble_2015,miles_observations_2020,suarez_ultracool_2022,beiler_first_2023, luhman_jwstnirspec_2023}, but the ability to see them in a single spectrum is groundbreaking. Of particular note is the absence of a resolved phosphine ($\mathrm{PH}_3$) feature in any of our objects, which is in direct contradiction to the predictions of atmospheric models with disequilibrium chemistry \citep[e.g.][]{meisner_exploring_2023, mukherjee_sonora_2023}, as has been previously noted in \citet{beiler_first_2023} and \citet{luhman_jwstnirspec_2023}.

With this large sample that spans the spectral type} regime, we can now see how these features evolve as a function of spectral type. We plot a spectral sequence in Figure \ref{fig:AbsorptionSequenceAbsNorm}, preferentially selecting objects with Spitzer colors and M$_{[4.5]}$ typical of the 20 pc population. There is a clear sequence of the flux decreasing shortward of 2 \mic{} and narrowing of the $J$ and $H$ bands, which we would expect given that this progression of features is what defines these spectral types. The near-infrared peaks are disappearing for the later-type objects as even the deeper layers probed by these atmospheric windows reach temperatures where we start to see the collapse of the Wien tail. The decreasing trend extends to 4 \mic{} and is also the case in the valley between the 5 and 10 \mic{} peak. This is caused by the blackbody peak moving redward for cooler objects and/or the increasing absorption from water and methane. 

Similarly, past 8 \mic{} we see a sequence in the ammonia features. The absorption in this portion of the spectrum is increasing in later spectral types, seen most clearly in the deepening of the ammonia $P$ and $R$ branches, and the $Q$-branch doublet. However, as these features deepen, the normalized flux actually increases as the blackbody peak moves closer to these wavelengths.

This leaves the 5 \mic{} peak, where there is almost no sequence. CO and CO$_2$ both have strong bands in this regime, and small changes in a variety of atmospheric parameters can have a large effect on these molecular bands. For one, the atmospheric composition, specifically the metallicity and the C/O ratio, affects the CO and CO$_2$ abundances since they are 2- and 3-metal molecules made of carbon and oxygen. Another factor in the lack of a clean sequence around the 5 \mic{} region is disequilibrium chemistry. At the higher temperatures of L and early T dwarfs, carbon dioxide and carbon monoxide are the dominant reservoirs of carbon over methane, which is why the earliest objects have strong CO and CO$_2$ bands. Yet even at the cooler temperatures of later spectral types where CH$_4$ is chemically favored, we still see significant amounts of CO and CO$_2$, as it is dredged from the deeper, warmer layers of the atmosphere into the photosphere faster than it can be converted into CH$_4$. Having so many parameters that can affect the CO and CO$_2$ abundances obfuscates any trends that might occur as a function of spectral type. Surface gravity also strongly impacts these species, as seen in \citet{zahnle_methane_2014}. At the temperatures of Y dwarfs, low-gravity objects will have enhanced CO and CO$_2$ absorption compared to high-gravity objects of the same temperature. 

One trend we do see in this wavelength range is the repression of the blue edge of the 5 \mic{} peak as we go to later spectral types, driven by the increasing abundance of methane. The increased abundance of methane with later spectral types is primarily tied to the decrease in effective temperature.

\section{Bolometric Luminosities and Effective Temperatures}\label{sec:lbolandteff}
By design, we are able to collect a substantial fraction of the light emitted by the objects in our sample due to the broad spectral coverage of JWST, resulting in precise bolometric fluxes. Calculating \fbol{}~($F_\mathrm{bol} = \int_0^\infty f_\lambda \, d\lambda$) requires constructing a complete spectral energy distribution from 0 to $\infty$ \mic{}. To do so, we linearly interpolate between zero flux at 0 \mic{} to the first data point of the NIRSpec spectrum, and then from the last data point through the JWST photometric points at 15.065, 17.987, and 20.795 \mic. Longward of this last photometric point, the spectral energy distribution can be approximated as a Rayleigh-Jeans tail, 
\begin{equation}
    f_{\lambda} = \frac{2ck_\mathrm{B}T}{\lambda^4} = \frac{C_\mathrm{RJ}}{\lambda^4},
\end{equation}
where $c$ is the speed of light, $k_\mathrm{B}$ is the Boltzmann constant, $T$ is temperature, and $C_\mathrm{RJ}$ is the proportionality constant. We solve for $C_\mathrm{RJ}$ using the F2100W flux density at 20.795 \mic, which gives us a functional form that allows us to extend the spectral energy distribution to $\infty$ \mic.

To calculate \fbol, we integrate this spectral energy distribution. The regions of linear interpolation and the Rayleigh-Jeans tail can be calculated analytically, while the spectrum is integrated numerically using Simpson's rule, which is reliable at the low resolution of our JWST spectra. We estimate the uncertainty of \fbol{} by generating one million spectral energy distributions from the uncertainty distributions on each data point. This includes randomly sampling the absolution calibration scaling factors derived from the Spitzer [4.5] and F1000W photometric points, allowing the NIRSpec and MIRI LRS spectra (and their uncertainties) to scale independently. We calculated \fbol{} for each of these million spectral energy distributions, with the mean and standard deviation of the resulting distributions reported in Table \ref{tbl:Teffs}. Our \fbol{} measurements have fractional uncertainties under 2\% (except for WISE J2102$-$44, which has a fraction uncertainty of 2.2\%), which is well within our 3\% goal.

We then calculate \lbol{} as $L_\mathrm{bol} = 4\pi d^2F_\mathrm{bol}$ using distances derived from the precise parallaxes listed in Table \ref{tbl:prop}. We report these bolometric luminosities in Table \ref{tbl:Teffs}. With such precise \fbol{} measurements, the parallax uncertainties becomes the dominant term in the bolometric luminosity uncertainties.

We could now calculate the effective temperature of our sample if our objects had known radii, since 
\begin{equation} \label{eqn:teff}
    T_\mathrm{eff} = \left( \frac{L_\mathrm{bol}}{4 \pi \sigma R^2} \right)^{\frac{1}{4}}.
\end{equation}
It is currently unfeasible to make direct measurements of the radii of the solivagant brown dwarfs that make up our sample. However, the radii of brown dwarfs are $\sim1\pm0.1$ $\mathcal{R}_\mathrm{J}^\mathrm{N}$—the nominal value for Jupiter’s equatorial radius of $7.1492 \times 10^7$ m \citep{mamajek_iau_2015}—across a wide range of ages and masses due to the competing effects of Coulomb repulsion and electron degeneracy \citep[e.g.][]{burrows_theory_2001}. We adopt this radius of 1 R$_\mathrm{Jup}$ for a nominal effective temperature $T_\mathrm{eff}^{R_\mathrm{Jup}}$, which we report in Table \ref{tbl:Teffs}.

\begin{deluxetable*}{llccccccc}
\tablecaption{Effective Temperature with Various Radius Assumptions\label{tbl:Teffs}}
\tablehead{
\colhead{Name} &
\colhead{SpT} &
\colhead{$F_\mathrm{bol}$} &
\colhead{$L_\mathrm{bol}$} &
\colhead{$\mathrm{log}(L/\mathcal{L}_\odot^N)^\tablenotemark{a}$ } &
\colhead{$T_\mathrm{eff}^{R_\mathrm{Jup}}$} &
\colhead{$T_\mathrm{eff}^{\mathrm{uni}}$} &
\colhead{$T_\mathrm{eff}^{\mathrm{B24}}$} \\
\colhead{} &
\colhead{} &
\colhead{(W m$^{-2}$)} &
\colhead{(W)} &
\colhead{} &
\colhead{(K)} &
\colhead{(K)} &
\colhead{(K)} 
}
\startdata
WISE J0247+37 & T8                      & $1.930\pm0.033\times 10^{-16}$ & $5.497\pm0.355\times 10^{20}$ & $-5.843\pm0.028$&  623 & $669^{+22}_{-35}$ & $627^{+35}_{-40}$ \\ 
WISEPA J0313+78 & T8.5                  & $5.578\pm0.078\times 10^{-16}$ & $3.625\pm0.156\times 10^{20}$ & $-6.024\pm0.019$&  562 & $598^{+17}_{-27}$ & $560^{+29}_{-31}$ \\ 
WISE J0359$-$54 & Y0                    & $6.952\pm0.124\times 10^{-17}$ & $1.536\pm0.088\times 10^{20}$ & $-6.397\pm0.025$&  453 &  $468^{+13}_{-23}$ & $443^{+23}_{-19}$ \\ 
WISE J0430+46 & T8                      & $1.781\pm0.033\times 10^{-16}$ & $2.310\pm0.144\times 10^{20}$ & $-6.219\pm0.027$&  502 & $525^{+16}_{-26}$ & $495^{+28}_{-23}$ \\ 
WISE J0535$-$75 & Y1                    & $9.170\pm0.153\times 10^{-17}$ & $2.326\pm0.139\times 10^{20}$ & $-6.216\pm0.026$&  503 & $526^{+16}_{-26}$ & $496^{+28}_{-23}$ \\ 
WISE J0734$-$71 & Y0                    & $8.568\pm0.154\times 10^{-17}$ & $1.846\pm0.091\times 10^{20}$ & $-6.317\pm0.022$&  475 & $493^{+14}_{-24}$ & $466^{+24}_{-20}$ \\ 
WISE J0825+28 & Y0.5                    & $1.786\pm0.029\times 10^{-16}$ & $9.168\pm0.293\times 10^{19}$ & $-6.621\pm0.014$&  398 & $406^{+10}_{-16}$ & $387^{+15}_{-15}$ \\ 
ULAS J1029+09 & T8                      & $3.316\pm0.053\times 10^{-16}$ & $8.435\pm0.438\times 10^{20}$ & $-5.657\pm0.023$&  694 & $756^{+23}_{-39}$ & $705^{+39}_{-49}$ \\ 
CWISEP J1047+54 & Y1                    & $3.886\pm0.063\times 10^{-17}$ & $1.002\pm0.145\times 10^{20}$ & $-6.582\pm0.063$&  407 & $415^{+20}_{-22}$ & $395^{+23}_{-21}$ \\ 
WISE J1206+84 & Y0                      & $1.154\pm0.020\times 10^{-16}$ & $1.927\pm0.100\times 10^{20}$ & $-6.298\pm0.023$&  480 & $499^{+14}_{-24}$ & $472^{+26}_{-20}$ \\ 
SDSSp J1346$-$00 & T6.5                 & $1.126\pm0.016\times 10^{-15}$ & $2.816\pm0.191\times 10^{21}$ & $-5.133\pm0.030$& 1011 &  $1044^{+26}_{-55}$& $976^{+58}_{-87}$ \\ 
WISEPC J1405+55 & Y0.5                  & $2.021\pm0.033\times 10^{-16}$ & $9.657\pm0.344\times 10^{19}$ & $-6.598\pm0.015$&  404 & $412^{+10}_{-17}$ & $392^{+16}_{-15}$ \\ 
CWISEP J1446$-$23 & Y1                  & $5.736\pm0.092\times 10^{-17}$ & $6.366\pm0.629\times 10^{19}$ & $-6.779\pm0.043$&  364 & $366^{+14}_{-14}$ & $351^{+16}_{-13}$ \\ 
WISE J1501$-$40 & T6                    & $6.993\pm0.110\times 10^{-16}$ & $1.579\pm0.101\times 10^{21}$ & $-5.385\pm0.028$&  811 & $897^{+27}_{-48}$ & $836^{+49}_{-68}$ \\ 
WISEPA J1541$-$22 & Y1                  & $2.671\pm0.045\times 10^{-16}$ & $1.147\pm0.033\times 10^{20}$ & $-6.523\pm0.013$&  421 & $432^{+11}_{-18}$ & $411^{+18}_{-17}$ \\ 
SDSS J1624+00 & T6                      & $1.595\pm0.022\times 10^{-15}$ & $2.265\pm0.066\times 10^{21}$ & $-5.228\pm0.013$&  888 & $987^{+22}_{-60}$ & $922^{+53}_{-78}$ \\ 
WISEPA J1959$-$33 & T8                  & $6.224\pm0.090\times 10^{-16}$ & $1.058\pm0.052\times 10^{21}$ & $-5.558\pm0.021$&  734 & $805^{+24}_{-43}$ & $750^{+42}_{-56}$ \\ 
WISEPC J2056+14 & Y0                    & $3.443\pm0.056\times 10^{-16}$ & $2.077\pm0.067\times 10^{20}$ & $-6.266\pm0.014$&  489 & $510^{+14}_{-25}$ & $481^{+26}_{-20}$ \\ 
WISE J2102$-$44 & T9                    & $2.434\pm0.054\times 10^{-16}$ & $3.372\pm0.157\times 10^{20}$ & $-6.055\pm0.020$&  552 & $585^{+17}_{-27}$ & $549^{+29}_{-30}$ \\ 
WISEA J2159$-$48 & T9                   & $1.371\pm0.027\times 10^{-16}$ & $3.004\pm0.221\times 10^{20}$ & $-6.105\pm0.032$&  536 & $565^{+20}_{-26}$ & $532^{+30}_{-28}$ \\ 
WISE J2209+27 & Y0                      & $1.449\pm0.025\times 10^{-16}$ & $6.621\pm0.207\times 10^{19}$ & $-6.762\pm0.014$&  367 & $371^{+9}_{-13}$ & $355^{+13}_{-11}$ \\ 
WISEA J2354+02 & Y1                     & $8.642\pm0.154\times 10^{-17}$ & $6.067\pm0.319\times 10^{19}$ & $-6.800\pm0.023$&  359 & $362^{+10}_{-12}$ & $347^{+13}_{-11}$ \\ 
WISEA J2018$-$74 $^\tablenotemark{b}$& T7 & $5.019\pm0.136\times 10^{-16}$& $8.677\pm0.458\times 10^{20}$& $-5.645\pm0.023$&    721 & $784^{+31}_{-50}$ & $733^{+47}_{-58}$\\
\enddata
\tablenotetext{a}{$\mathcal{L}_\odot^N$ is the Sun's nominal luminosity of 3.828 $\times$ 10$^{26}$ W \citep{mamajek_iau_2015}.}
\tablenotetext{b}{This object is lacking NIRSpec data, which is replaced in the spectral energy distribution by a near-infrared spectrum (Table \ref{tbl:nir}) and W1 and [4.5] photometry. The effective temperatures have a 23$\pm$8 K correction applied, as detailed in Section \ref{sec:Varients}.}
\end{deluxetable*}

A more rigorous estimate of our objects' radii can be made using the technique of \citet{saumon_molecular_2000}, which relies on evolutionary models. Evolutionary models use our current understanding of substellar atmospheric and interior physics to predict the relations between the fundamental parameters of brown dwarfs including the age, radius, and luminosity. With these models, we can estimate the radius distributions for our objects from our observed luminosity and an age distribution.

Our objects have no age constraints due to not being a part of young moving groups or binary systems, so we make the reasonable assumption of a field age distribution. We report temperatures assuming two age distributions: 1) a uniform 1$-$10 Gyr distribution as preferred by \citet{kirkpatrick_initial_2024} and 2) an exponential distribution ($p_{\mathrm{age}}(t)=e^{-\beta t}$) where $t\in[0,10]$ Gyr and $\beta=0.44$, as preferred by \citet{best_volume-limited_2024}. This exponential distribution is skewed younger, which results in larger radii and cooler temperatures due to brown dwarfs contracting and cooling as they age.

\begin{figure*}
\includegraphics[width=0.7\textwidth]{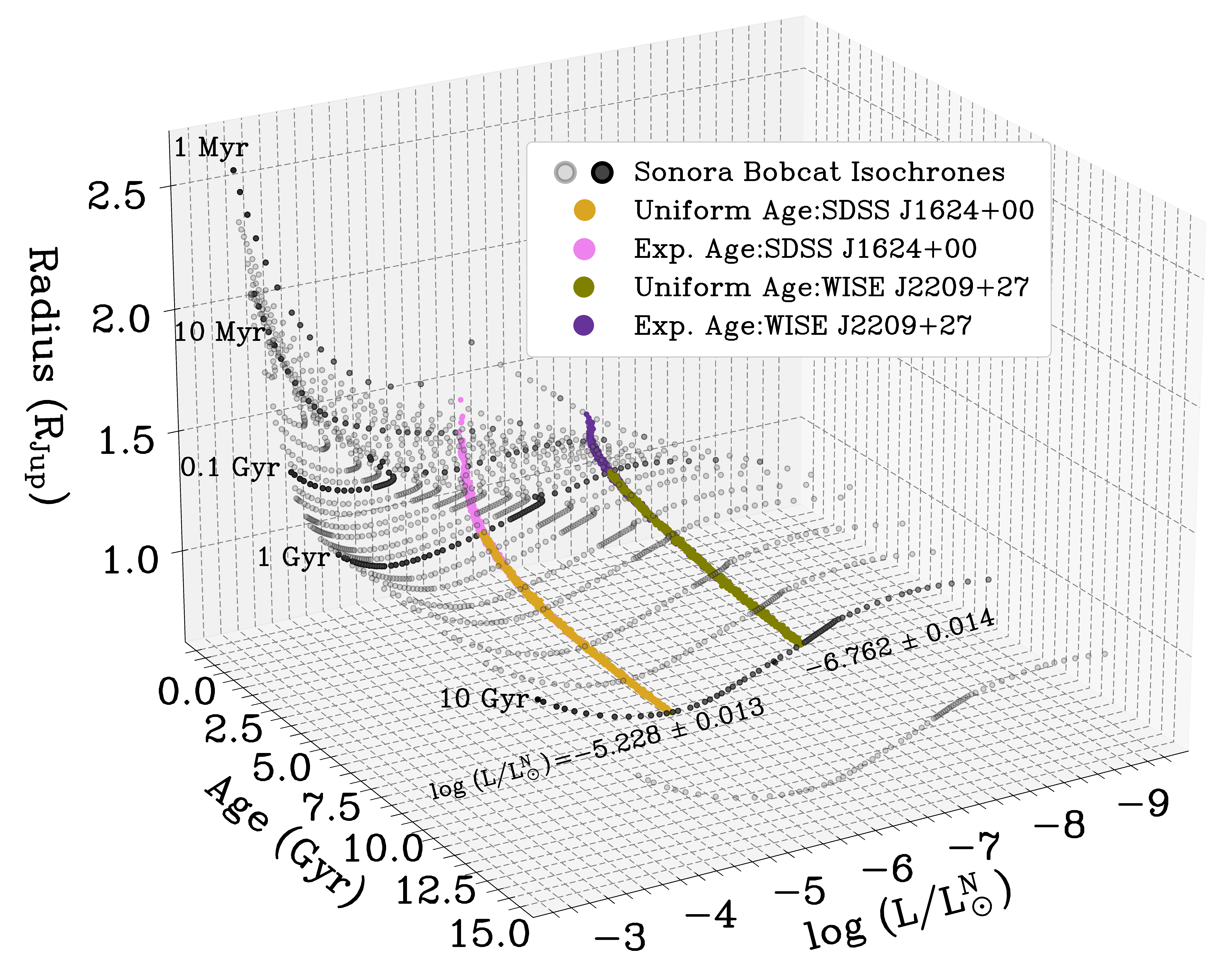}
\centering
\caption{The Sonora Bobcat solar-metallicity evolutionary model isochrones (grey and black) plotted in luminosity-age-radius space, which we use to estimate the radii of our objects. The 0.001, 0.01, 0.1, 1, and 10 Gyr isochrones are plotted in black, and the model isochrones are more finely sampled at low ages and more sparsely sampled at older ages. Also shown are drawn radii points of two objects: SDSS J1624+00 (uniform age distribution in yellow, exponential age distribution in pink), one of our more luminous objects, and WISE J2209+27 (uniform in green, exponential in purple), one of our dimmer objects. The exponential age distribution does extend to the 10 Gyr, but it is hidden by the uniform age distribution points. There is a local minima in the evolutionary model grid at $\mathrm{log}(L/\mathcal{L}_\odot^N)$ $\sim$ $-$4.5 that results in the more luminous objects in our sample having smaller radii than our dimmer objects. For each distribution, the plotted points are only 1000 of the one million randomly selected points, as this is sufficient to show the coverage of the distributions.}
\label{fig:Evogrid}
\end{figure*}

For this work, we draw one million ages and luminosities from their respective distributions, and use the Sonora Bobcat solar metallicity evolutionary model \citep{marley_sonora_2021}, part of new generation of self-consistent atmospheric and evolutionary models that include updated opacities and atmospheric chemistry. We perform a linear interpolation across the luminosity-age-radius grid for pairs of ages and luminosities to arrive at a radius distribution. The evolutionary model isochrones are plotted in Figure \ref{fig:Evogrid}, more finely sampled at low ages and more sparsely at older ages. Also plotted are the drawn radii for two example objects: SDSS J1624+00, one of our more luminous objects, and WISE J2209+27, one of our dimmer objects. Counter-intuitively, for our sample the more luminous objects have a smaller radius, as the evolutionary model isochrones have a local radius minimum at $\mathrm{log}(L/\mathcal{L}_\odot^N)$ $\sim$ $-$4.5.

We translate the radius and luminosity distributions into a temperature distribution using Eq. \ref{eqn:teff}. We summarize the temperature distributions with their mean values and uncertainties that contain 34.1\% of the distribution above and below the mean respectively, which we report in Table \ref{tbl:Teffs}. Our hottest object is SDSSp J1346$-$00 at $\sim1000$ K, and our coldest objects are WISEA J2354+02 and CWISEP J1446$-$23 at $\sim360$ K.

We also tested uniform age distributions that extended down to 0.5 Gyr to see how this would vary \teff. This resulted in 1$-$5 K colder temperatures, with larger differences for hotter objects. Uncertainties also increase by 1$-$12 K, with the larger increases once again coming from the hotter objects. The changes are a result of including even younger objects, which extends the tail of the radius distributions to larger radii. The fractional change in the mean \teff{} is under 1\% for all objects, so we feel confident moving forward with our uniform 1$-$10 Gyr age distribution.

\begin{figure*}
\includegraphics[width=\textwidth]{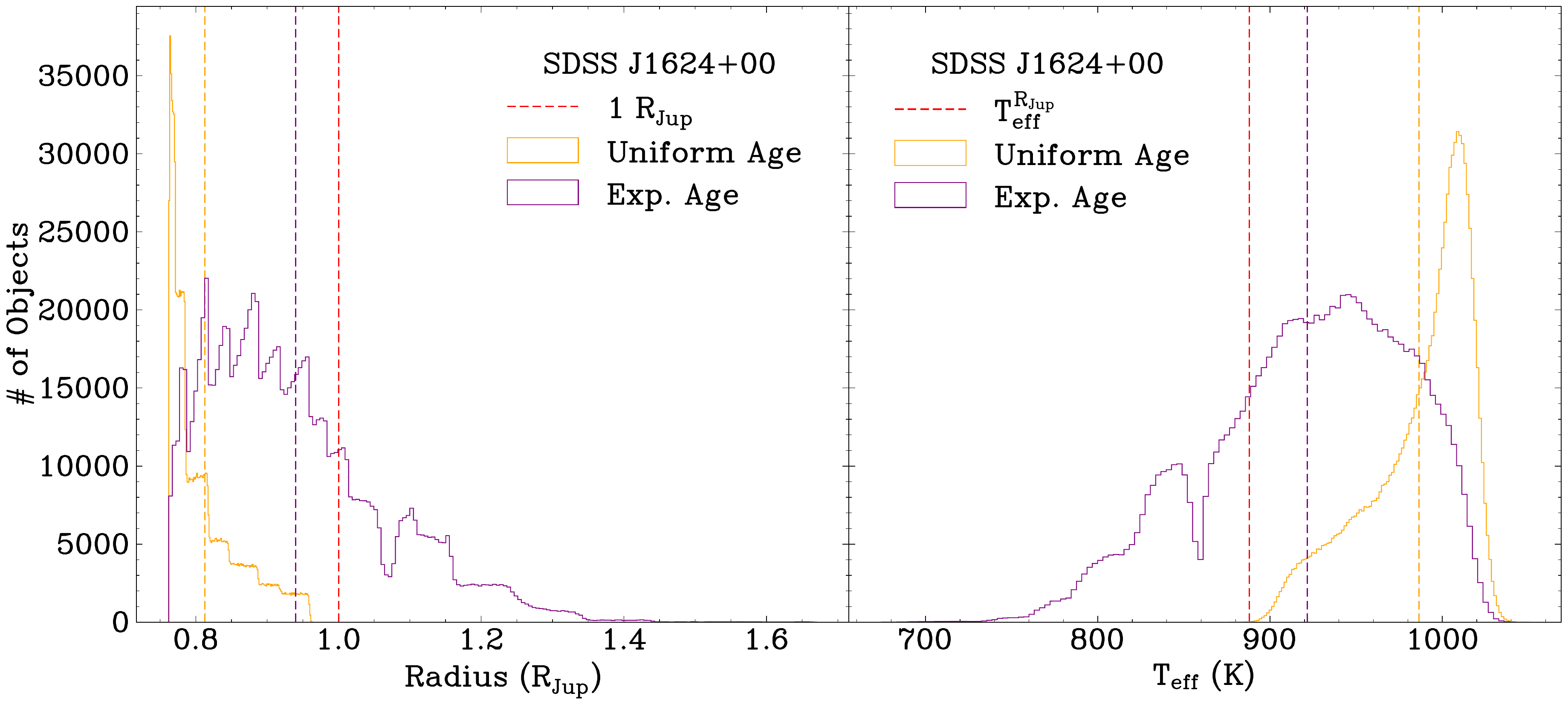}
\centering
\caption{Left: the distribution of a million radii drawn from Sonora Bobcat solar metallicity models using the luminosity of SDSS J1624+00 ($2.265\pm0.064 \times 10^{21}$ W) and either a 1$-$10 Gyr uniform age distribution (gold) or a negative exponential age distribution (purple, described more fully in Section \ref{sec:lbolandteff}). The mean radius for each distribution is plotted as dashed vertical line, as well as 1 R$_\mathrm{Jup}$. The bump from 1.07$-$1.25 R$_\mathrm{Jup}$ is due to deuterium burning which occurs at $\sim$0.4 Gyr for this luminosity. The different age distributions have a large disparity in their predicted radii with the young ages of the exponential age distribution allowing for radii up to 1.6 R$_\mathrm{Jup}$. Right: the effective temperature distributions calculated using the radius distributions on the left and the luminosity. The mean effective temperature for each distribution is plotted as dashed vertical line, as well as the nominal effective temperature, which assumes R = 1 R$_\mathrm{Jup}$. The bump from 820$-$850 K is tied to deuterium burning. There is a significant difference in the mean \teff{} values of the two age assumptions, but an even larger difference in the broadness of the \teff{} distribution.} \label{fig:RadiiDistHot}
\end{figure*}

\begin{figure*}
\includegraphics[width=\textwidth]{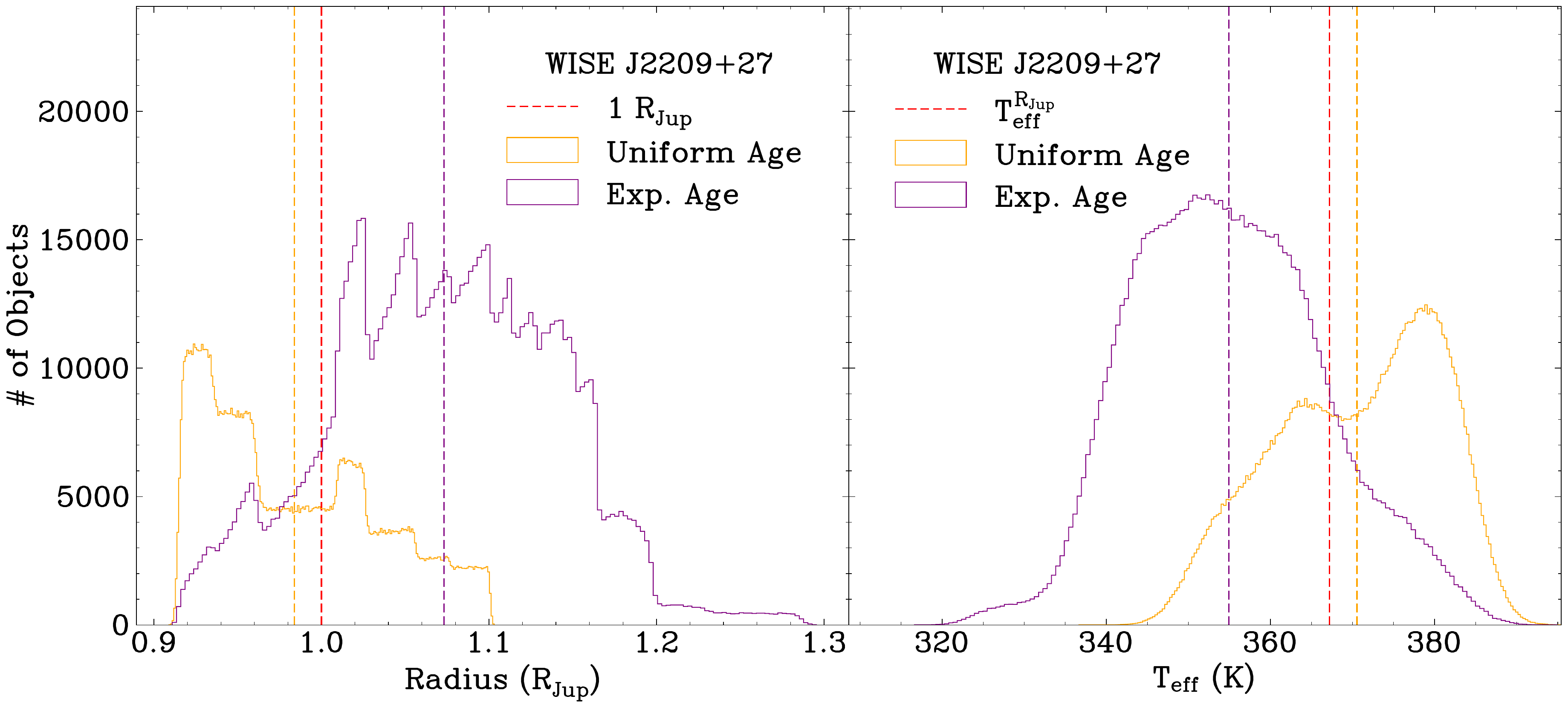}
\centering
\caption{The same as Figure \ref{fig:RadiiDistHot}, but for WISE J2209+27 ($6.621\pm0.208 \times 10^{19}$ W). Here the deuterium burning bump is from 1.02$-$1.06 R$_\mathrm{Jup}$ and 360$-$370 K, and occurs at $\sim$4 Gyr for this luminosity. For the colder objects, the two age assumptions have a similar spread in bulk of their distributions, though the exponential age distribution results in longer tails, especially at the young age, high radius end.} \label{fig:RadiiDistCold}
\end{figure*}

As examples, we plot the radius and temperature distributions SDSS J1624+00 and WISE J2209+27, in Figures \ref{fig:RadiiDistHot} and \ref{fig:RadiiDistCold}. Note that due to their inverse relation, a larger radius results in cooler temperature, and vice versa. We see step-like and spiky features in the uniform age and exponential age radius distributions respectively due to our choice of linear interpolation and the sparseness of the evolutionary model age sampling at older ages. These two factors result in large age ranges between isochrones where the interpolated surface has a constant slope and therefore results in a roughly constant probability density for a given range of radii. A cubic interpolation was attempted, but this results in both oscillatory artifacts and false peaks in the interpolated surface, especially at younger ages, that have a much stronger impact on the radius distribution than the linear interpolations artifacts. The linear interpolation, while not creating a smooth radius distribution, is sufficiently accurate for our work.

In comparing the temperature distributions for these two objects, we see that for the hotter object, the exponential age distribution results in a much wider range of radii and therefore also a wider range of effective temperatures. At the hot end, this distribution results in $\sim$7\% uncertainty, but falls to $\sim$3\% for the colder objects. On the other hand, the uniform age distribution is consistently $\sim$3\% across our sample.

No matter our choice of age distribution, the radius estimation ends up dominating the effective temperature uncertainty for all objects (with the exception of CWISEP J1446$-$23 and CWISEP J1047+54 whose higher parallax uncertainties result in $\sigma_{L_\mathrm{bol}}$ equally contributing to $\sigma_{T_\mathrm{eff}}$). A marginal improvement in the precision of \teff{} could be made with further astrometric follow-up or by expanding the spectral energy distributions. However, even if we knew our bolometric luminosity to infinite precision, we would still be unable to lower our fractional \teff{} uncertainty below 2.5\% without stronger age constraints or direct radius measurements.

Previous effective temperature estimates for these objects can be found in \citet{kirkpatrick_field_2021}, assigned either from measured values in \citet{filippazzo_fundamental_2015} or the \teff{} versus M$_\mathrm{H}$ or spectral type relations. With the broad spectral coverage of JWST, we are able to significantly reduce the uncertainly on \teff, in some cases by a factor of 7. The previous estimates were accurate, despite their low precision, as all but two of our objects fall within $1\sigma$ of the \citet{kirkpatrick_field_2021} temperatures. As for systematic differences, the uniform age distribution tends to result in higher \teff{} than those of \citet{kirkpatrick_field_2021}, while the exponential age distribution results in a slightly lower temperature in comparison. 

\begin{figure*}
\includegraphics[width=\textwidth]{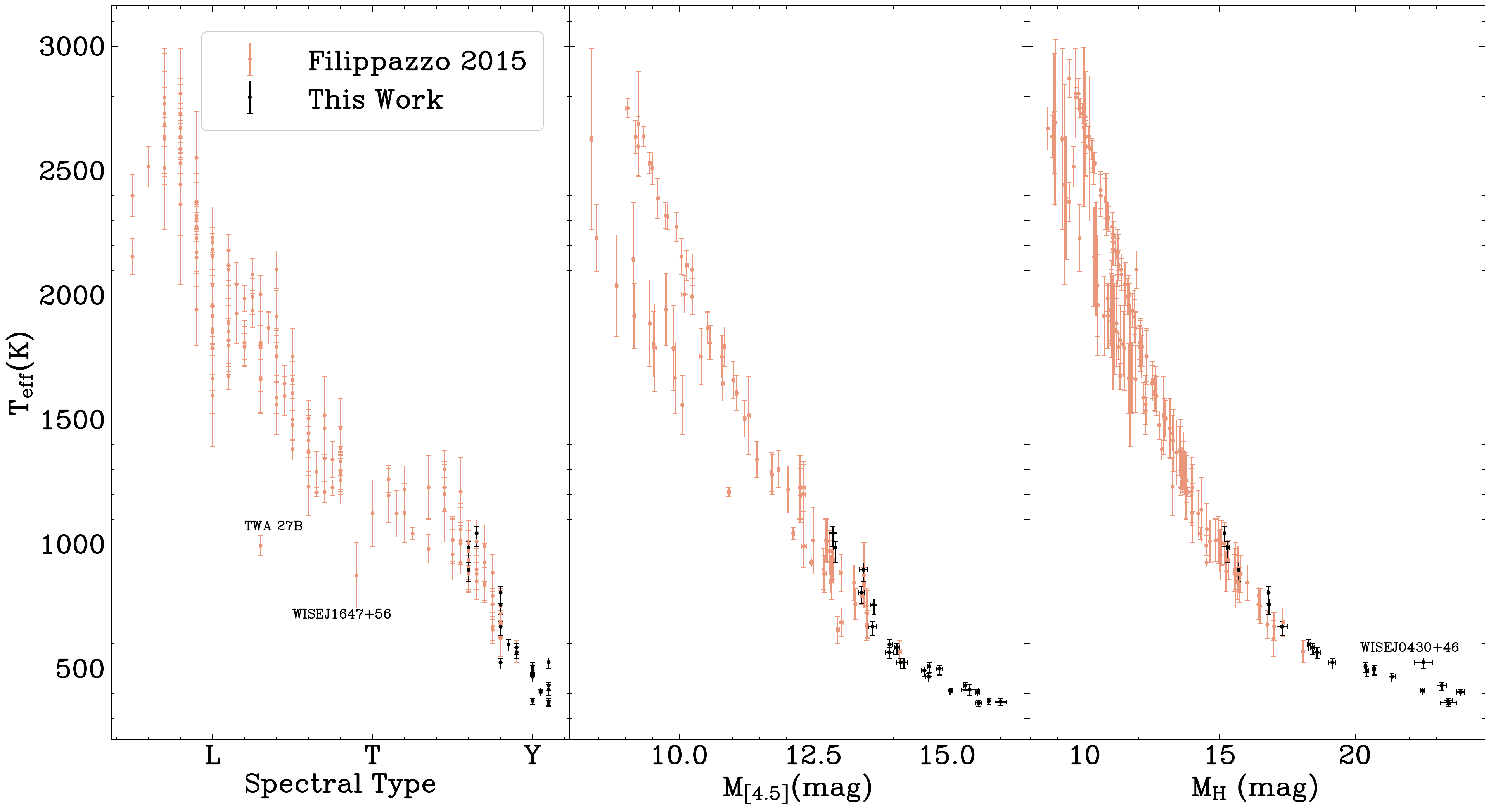}
\centering
\caption{Our effective temperatures (black) and effective temperatures from \citet{filippazzo_fundamental_2015} (orange) plotted as a function of spectral type, [4.5] absolute magnitude, and $H$-band absolute magnitude (where available). The samples overlap over a small range of parameters (including two shared objects), and where this happens our objects tend to have hotter effective temperatures. Objects that fall outside the population trends are labelled.} \label{fig:FiliComp}
\end{figure*}

The temperatures in \citet{kirkpatrick_field_2021} are largely derived from the previous work of \citet{filippazzo_fundamental_2015}, who constructed spectral energy distributions and calculated \teff{} for a large sample of late M to T dwarfs using a technique similar to ours. We constructed our sample to overlap in parameter space with \citet{filippazzo_fundamental_2015} and also included two objects from their sample, allowing us to compare effective temperature values. Figure \ref{fig:FiliComp} shows the \citet{filippazzo_fundamental_2015} temperatures and our uniform age distribution temperatures as a function of spectral type, M$_{[4.5]}$, and M$_{H}$. CWISEP J1446$-$23 and CWISEP J1047+54 do not have $H$-band photometry, and so are not included in this plot.

Looking at where the two samples overlap, we see that our \teff{} values tend to the hotter side of the \citet{filippazzo_fundamental_2015} distribution. For two overlapping objects SDSS J1624+00 and SDSSp J1346$-$00, our effective temperatures are 51 and 33 K hotter, respectively. This is true, even though the Filippazzo luminosities are brighter. We measure $\mathrm{log}(L/\mathcal{L}_\odot^N)$ as $-$5.227$\pm$0.013 and $-$5.133$\pm$0.030 respectively, compared to Filippazzo's measurements of $-$5.19$\pm$0.01 and $-$5.06$\pm$0.03. This means that the difference must be in the assumed radii. 

\citeauthor{filippazzo_fundamental_2015} calculated their radii by assuming an age of 0.5$-$10 Gyr for both objects, and using the evolutionary models from \citet{saumon_evolution_2008} and \citet{baraffe_evolutionary_2003} arrived at a range of possible radii, which they reported as $0.94\pm\sim0.16$ $R_\mathrm{Jup}$, compared to our value of $\sim0.81\pm0.05$ $R_\mathrm{Jup}$. The difference primarily arises from \citeauthor{filippazzo_fundamental_2015} selecting a radius \textit{range} from the evolutionary models, which creates a bias towards the larger radii at the tail of the distribution, while we draw a \textit{distribution} from the evolutionary models, which favors the smaller radii. This explains why our effective temperatures are systematically hotter than those of \citet{filippazzo_fundamental_2015}. The different choices of evolutionary models compounds this effect, as when we draw radii for these two objects from the \citet{saumon_evolution_2008} hybrid evolutionary models, we find our radii increases by 0.02 $R_\mathrm{Jup}$. Similarly, our choice of 1$-$10 Gyr shifts our radii distribution smaller, but this is also a secondary effect, as even if we assumed a uniform age of 0.1$-$10 Gyr, our mean radii would only increase by 0.02 $R_\mathrm{Jup}$. Both choices impact the reported radii, but even combined they cannot account for the 0.13 $R_\mathrm{Jup}$ difference. 

The brighter luminosities reported by \citeauthor{filippazzo_fundamental_2015} are also interesting in and of themselves. Updated parallaxes have remained within 1\% of the old values, so differences in distance are not the culprit, and both of these objects had coverage out to 12 \mic through WISE photometry or Spitzer InfraRed Spectrograph (IRS) spectra, so an overestimated Rayleigh-Jeans tail is not the issue. The likely cause of the discrepancy is the region from 2$-$6 \mic, which \citet{filippazzo_fundamental_2015} only sampled with Spitzer and WISE photometry. The linear interpolation through these points, especially across deep absorption features, could easily result in the luminosity difference we are seeing in these objects. The systematically brighter [3.6] magnitudes that we noted in Section \ref{subsec:PipeCali} would also exacerbate this issue. This shows once again how essential a complete spectral energy distribution is to calculating the effective temperatures of these cold objects.

\section{Incomplete Spectral Energy Distributions}\label{sec:Varients}
Few substellar objects will have their full spectral energy distributions observed with JWST, so it is useful to see how the effective temperatures computed using incomplete spectral energy distributions compare to our reported values. We look at six incomplete spectral energy distribution variations, which are constructed from:

1. JWST NIRSpec and MIRI LRS spectra,

2. JWST NIRSpec spectra,

3. JWST NIRSpec spectra and all four MIRI photometric points,

4. Previous near-infrared spectra, W1, W2, and [4.5] photometric points, and JWST MIRI LRS spectra,

5. Previous near-infrared spectra, W1, W2, and [4.5] photometric points, JWST MIRI LRS spectra, and our MIRI photometric points,

6. Previous near-infrared spectra and W1, W2, and [4.5] photometric points.

When the NIRSpec spectrum is not included, we replaced it with previous ground or Hubble observations, and Spitzer [4.5] and WISE W1 and W2 photometry when available, as these observations are nearly ubiquitous for cold objects. The references for the near-infrared spectra are found in Table \ref{tbl:nir}. The Spitzer [3.6] photometric point was omitted due to the systematic offset we noted in Section \ref{subsec:PipeCali}, and points are also omitted in cases where they overlap with a spectrum (such as the case for the W2 point when MIRI LRS is included). A handful of our objects do have Spitzer IRS and M band spectra, but we chose to not include these in our variant spectral energy distributions to make the results more universally applicable. 

When both NIRSpec and MIRI LRS are included, they are merged as before, but when only one is included we trim the spectrum to eliminate low signal-to-noise pixels or detector edge effects. For the NIRSpec data, we trim the spectrum at the first point past 4 \mic{} with a signal-to-noise ratio below 10 (cut-off wavelengths range from 4.94$-$5.29 \mic). The MIRI LRS data nominally covers down to 5 \mic, but we find that the spectra are reliable only down to 4.5 \mic, so we make our cut there.

\begin{deluxetable}{lr}
\tablecaption{Near-Infrared Spectral Reference\label{tbl:nir}}
\tablehead{
\colhead{Object Name} &
\colhead{Reference} 
}
\startdata
WISE J0247+37  & \citet{mace_study_2013}\\
WISEPA J0313+78& \citet{kirkpatrick_first_2011}\\
WISE J0359$-$54  & \citet{kirkpatrick_further_2012}\\
WISE J0430+46  & \citet{mace_study_2013}\\
WISE J0535$-$75  & \citet{schneider_hubble_2015}\\
WISE J0734$-$71  & \citet{kirkpatrick_further_2012}\\
WISE J0825+28  & \citet{schneider_hubble_2015}\\
ULAS J1029+09  & \citet{thompson_nearby_2013}\\
WISE J1206+84  &\citet{schneider_hubble_2015}\\
SDSSP J1346$-$00 &\citet{burgasser_unified_2006}\\
WISEPC1405+55 & \citet{schneider_hubble_2015}\\
WISE J1501$-$40  & \citet{tinney_new_2018}\\
WISEPA J1541$-$22& \citet{cushing_discovery_2011}\\
SDSS J1624+00  & \citet{burgasser_unified_2006}\\
WISEPA J1959$-$33& \citet{zhang_uniform_2021}\\
WISEA J2018$-$74 & \citet{burgasser_fire_2011}\\
WISEPC J2056+14& \citet{cushing_discovery_2011}\\
WISE J2102$-$44  & \citet{tinney_new_2018}\\
WISEA J2159$-$48  & \citet{tinney_new_2018}\\
WISE J2209+27  & \citet{schneider_hubble_2015}\\
WISEA J2354+02 & \citet{schneider_hubble_2015}\\
\enddata
\end{deluxetable}

\begin{figure*}
\includegraphics[width=\textwidth]{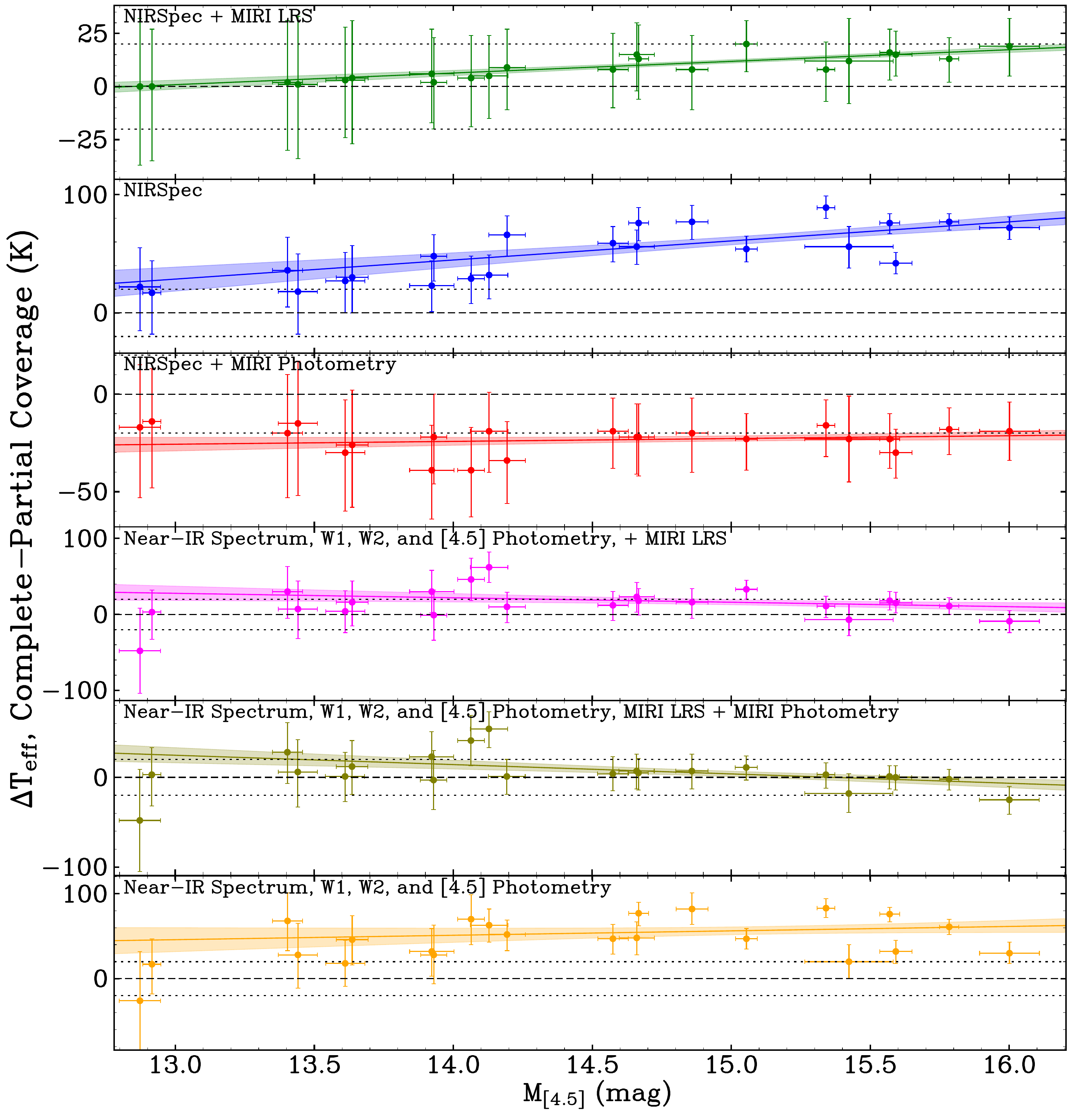}
\centering
\caption{The difference in effective temperatures derived from various partial spectral energy distributions and those from the full distribution as a function of the absolute [4.5] magnitude. For ease of comparison, we include dashed lines at 0 K and dotted lines at $\pm$20 K. Linear fits are shown for each data set (solid line), with the covariance errors plotted as the shaded region. More detail on how we build the spectral energy distributions can be found in Section \ref{sec:Varients}.} \label{fig:VarientTeffs}
\end{figure*}

In building these incomplete spectral energy distributions, we adopt a similar strategy as our full one. We linearly interpolate to zero flux at zero microns to the bluest point, attached a Rayleigh-Jeans tail to the reddest point, and in between, linearly interpolate through photometric points. These regions are all integrated analytically, whereas the spectra are integrated numerically using Simpson's Rule. We generate 100,000 spectral energy distributions (high-resolution near-infrared data makes generating one million difficult) based on the uncertainties on each data point from which we calculate \fbol, from which we calculate \lbol{} with the distance. For this comparison, since we only care about the difference in temperature between the complete and partial spectral energy distributions, we only calculate radii and temperatures based on the uniform age distribution. We plot the difference between the effective temperatures from the complete and partial spectral energy distributions (\teff$^\mathrm{full}-$\teff$^\mathrm{partial}$) in Figure \ref{fig:VarientTeffs}, with each panel showing a different partial spectral energy distribution. This is done as a function of [4.5] absolute magnitude, which roughly corresponds to spectral type. 

The only partial distribution that results in consistently hotter effective temperatures is the one that includes MIRI photometric points but not the MIRI LRS spectra, which is expected as linearly interpolating from the 5 \mic{} peak to the 10 \mic{} photometric point significantly overestimates the flux in this regime. This partial distribution also show almost no dependence on [4.5] absolute magnitude (fits listed in Table \ref{tbl:VarientTeffs}).

\begin{deluxetable}{cccccc}
\tablecaption{Linear Fits and Covariance Matricies of Difference in Temperatures Between Complete and Various Partial Spectral Energy Distributions \label{tbl:VarientTeffs}}
\tablehead{
\colhead{NIRSpec} &
\colhead{LRS} &
\colhead{Phot.} &
\colhead{$c_1$} &
\colhead{$c_0$} &
\colhead{$Cov$}
}
\startdata
X  & X  & -- &  5.5 &$-$70.4 & $\begin{pmatrix}  1.0 & -15.1 \\ -15.1 & 226.2 \end{pmatrix}$  \\
X  & -- & -- & 16.1 &$-$181.3& $\begin{pmatrix}  19.9& -301.9 \\ -301.9 & 4585.8 \end{pmatrix}$  \\
X  & -- & X  &  1.5 &$-$44.6 & $\begin{pmatrix}  2.8 & -41.1 \\ -41.1 & 616.2 \end{pmatrix}$  \\
-- & X  & -- & $-$5.9 &104.0 & $\begin{pmatrix}  18.9& -284.5 \\ -284.5 & 4286.3 \end{pmatrix}$  \\
-- & X  & X  &$-$10.4 &159.9 & $\begin{pmatrix}  15.6 & -234.2 \\ -234.2 & 3521.2 \end{pmatrix}$  \\
-- & -- & -- &  5.2 &$-$21.6 & $\begin{pmatrix}  39.5 & -597.3 \\ -597.3 & 9052.9 \end{pmatrix}$  \\
\enddata
\end{deluxetable}

The strongest dependence on [4.5] absolute magnitude comes from the spectral energy distribution with only the NIRSpec spectrum. For the hottest objects in our sample, NIRSpec plus a Rayleigh-Jeans tail mostly replicates the flux, but as we look at cooler objects and the peak of the blackbody curve moves redward, more and more flux is coming out at wavelengths not probed by NIRSpec. This $\sim$75 K offset for the coldest objects is one of the primary reasons why MIRI LRS data is so important for understanding Y dwarf atmospheres. The MIRI photometry is not quite as essential, but without it the coldest objects would be systematically off by $\sim1\sigma$ ($\sim$20 K).

We have particular interest in the partial spectral energy distribution which includes the MIRI data but not the NIRSpec spectrum, as this is the data we have available for WISEPA J2018$-$74. From this partial spectral energy distribution we calculate \fbol{} and \lbol{} (reported in Table \ref{tbl:Teffs}), which results in an effective temperature of 761$^{+23}_{-42}$ K for our uniform age distribution, and 710 $^{+39}_{-50}$ K for our exponential distribution. WISEPA J2018$-$74 has a [4.5] absolute magnitude of 13.15$\pm$0.05 which, given the fits listed in Table \ref{tbl:VarientTeffs}, requires a temperature correction of 23 $\pm$ 8 K. We therefore report the effective temperature of WISEPA J2018$-$74 as 784$^{+31}_{-50}$ K for our uniform age distribution, and 733 $^{+47}_{-58}$ K for our exponential distribution. This is the reported temperature in Table \ref{tbl:Teffs}.

\section{Evolution of Absorption Features with Effective Temperature}\label{sec:specEvol}
\begin{figure*}
\includegraphics[width=\textwidth]{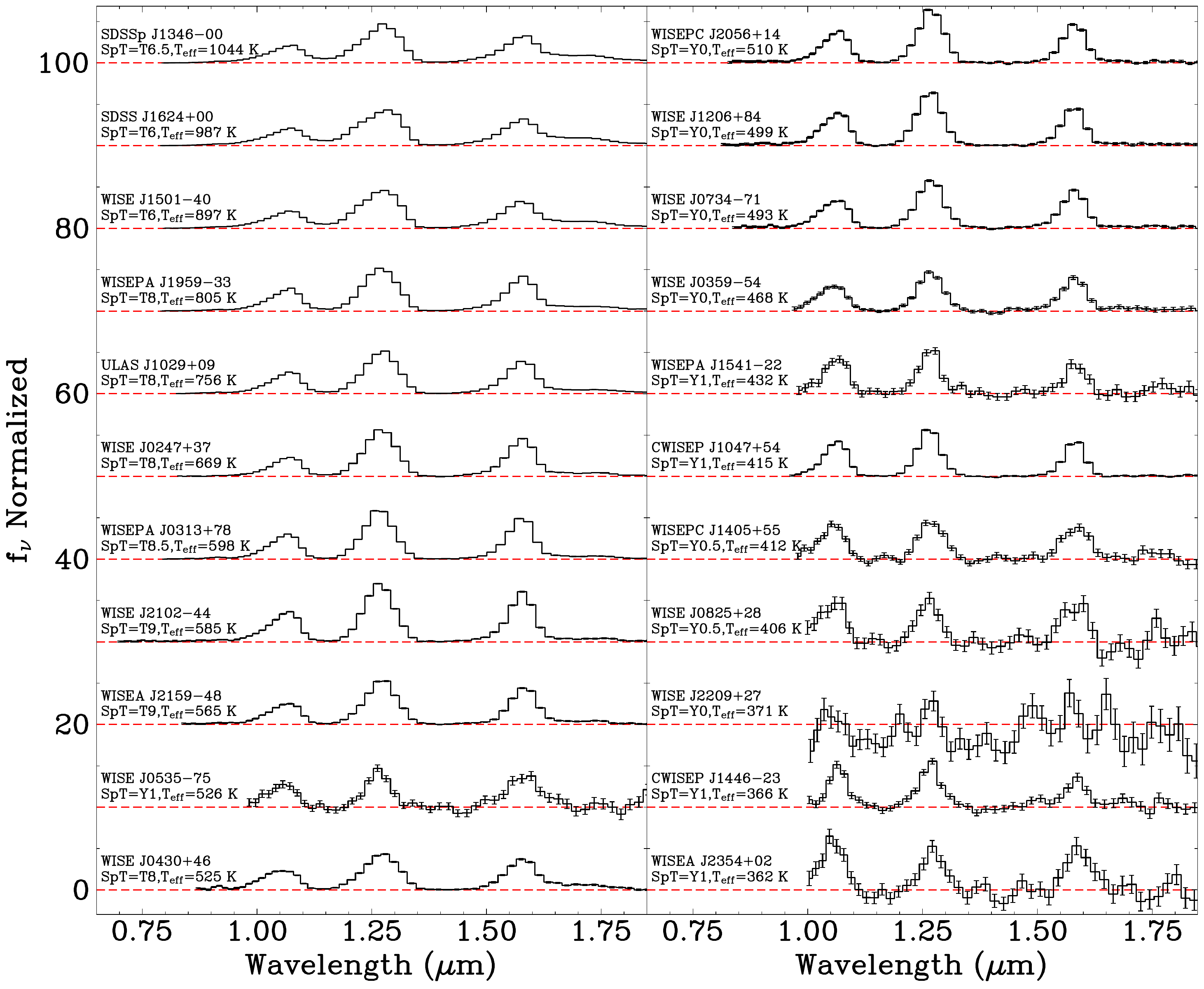}
\centering
\caption{Our sample of objects, sorted by effective temperature, zoomed in on the near-infrared portion of the spectra. Each object is normalized across this shown portion of the spectra, and offset to the red dashed line.} \label{fig:AllLadderNearIR}
\end{figure*}

\begin{figure*}
\includegraphics[width=\textwidth]{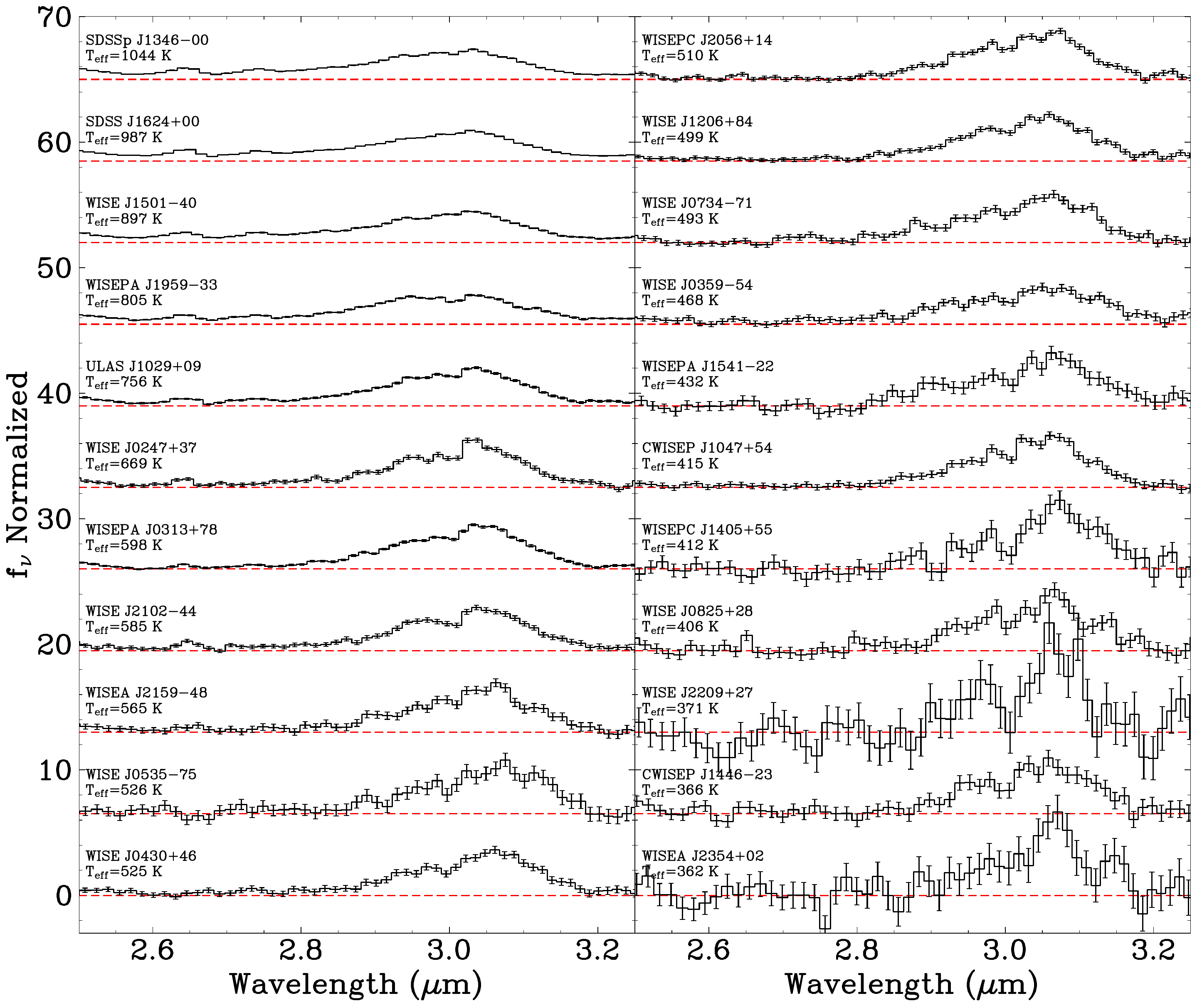}
\centering
\caption{Same as Figure \ref{fig:AllLadderNearIR}, but zoomed in on the 3 \mic{} ammonia feature.} \label{fig:AllLadder3mic}
\end{figure*}

\begin{figure*}
\includegraphics[width=\textwidth]{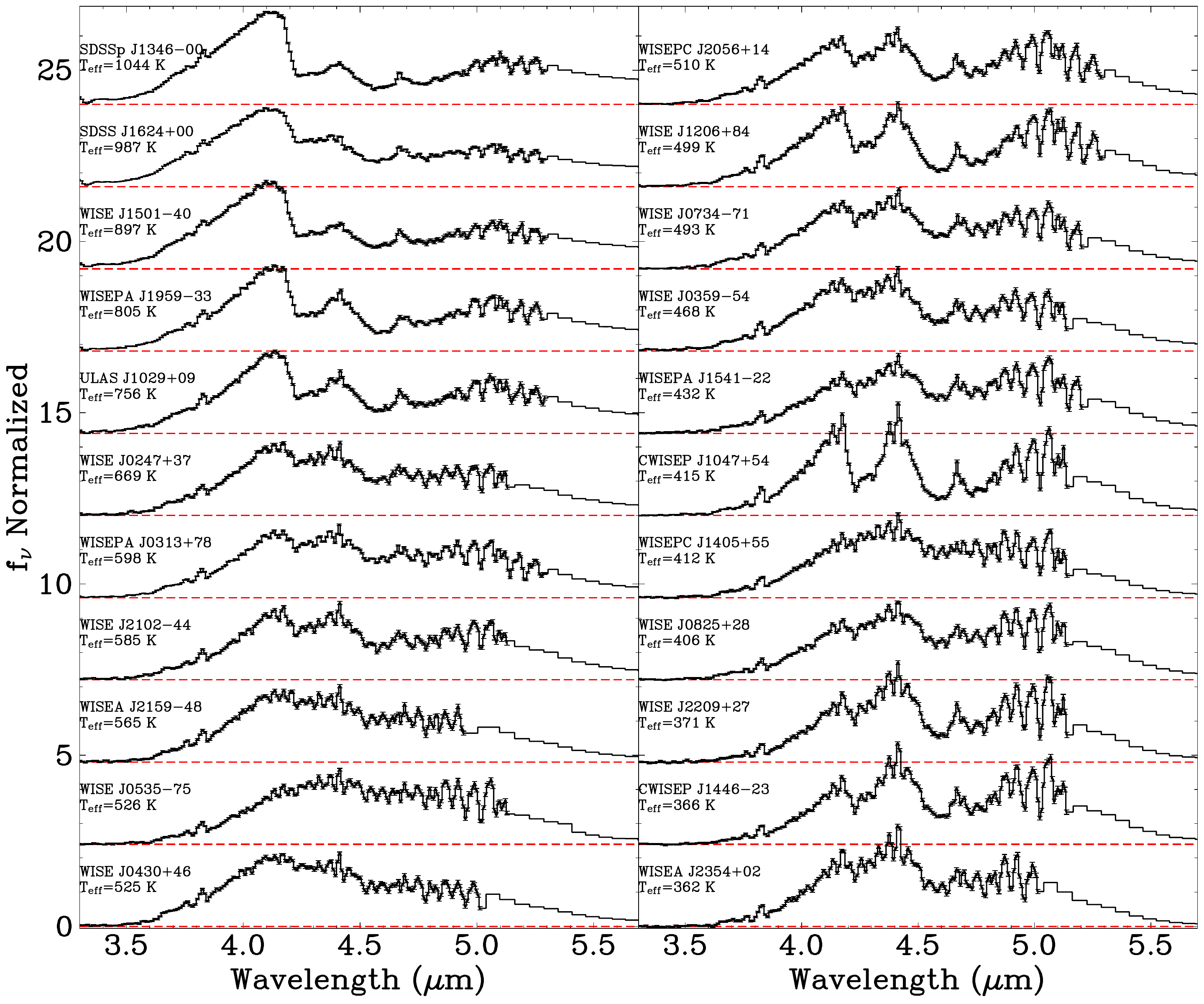}
\centering
\caption{Same as Figure \ref{fig:AllLadderNearIR}, but zoomed in on the 5 \mic{} peak.} \label{fig:AllLadder5mic}
\end{figure*}

\begin{figure*}
\includegraphics[width=\textwidth]{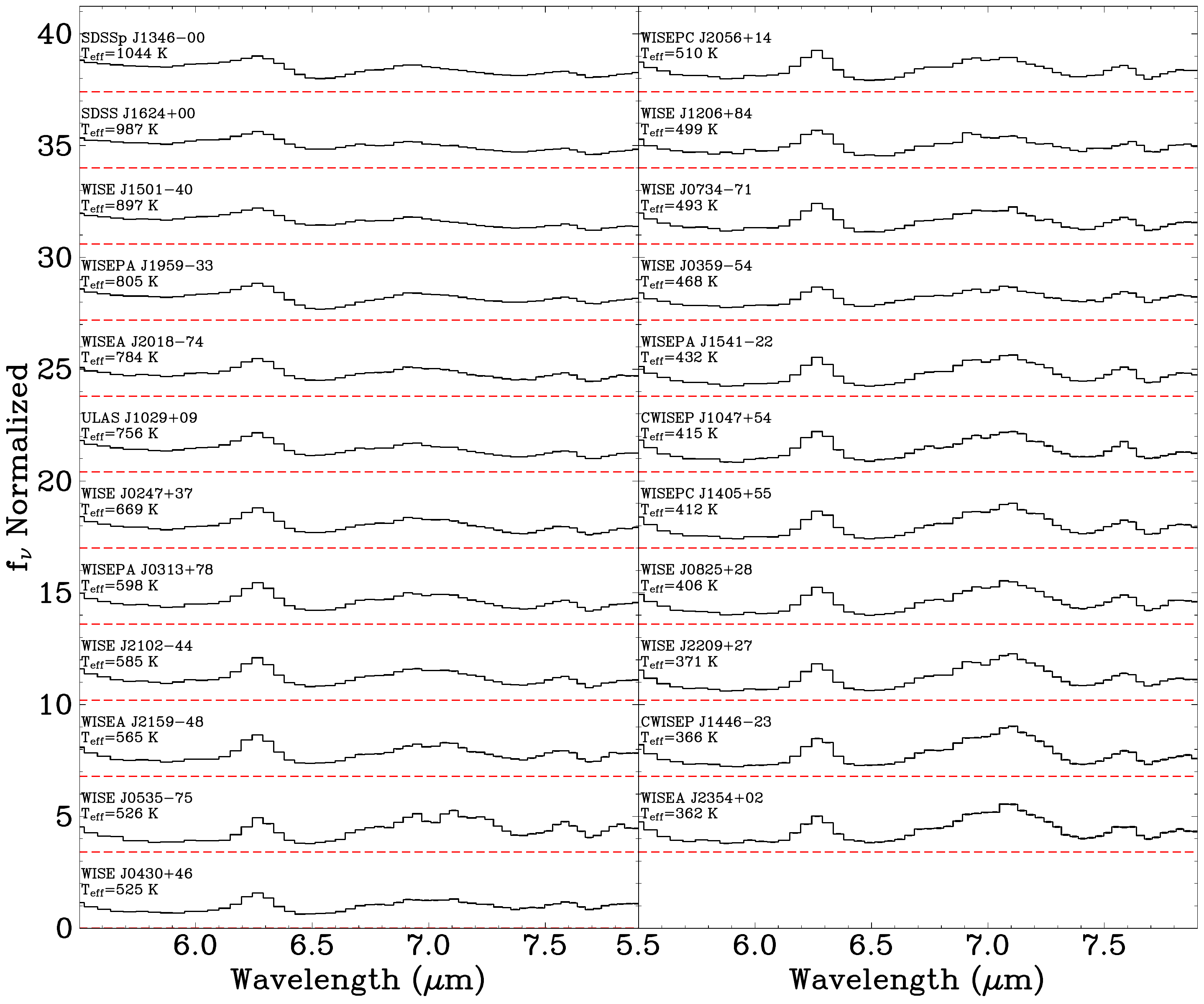}
\centering
\caption{Same as Figure \ref{fig:AllLadderNearIR}, but zoomed in on the valley between the 5 and 10 \mic{} peaks.} \label{fig:AllLadderValley}
\end{figure*}

\begin{figure*}
\includegraphics[width=\textwidth]{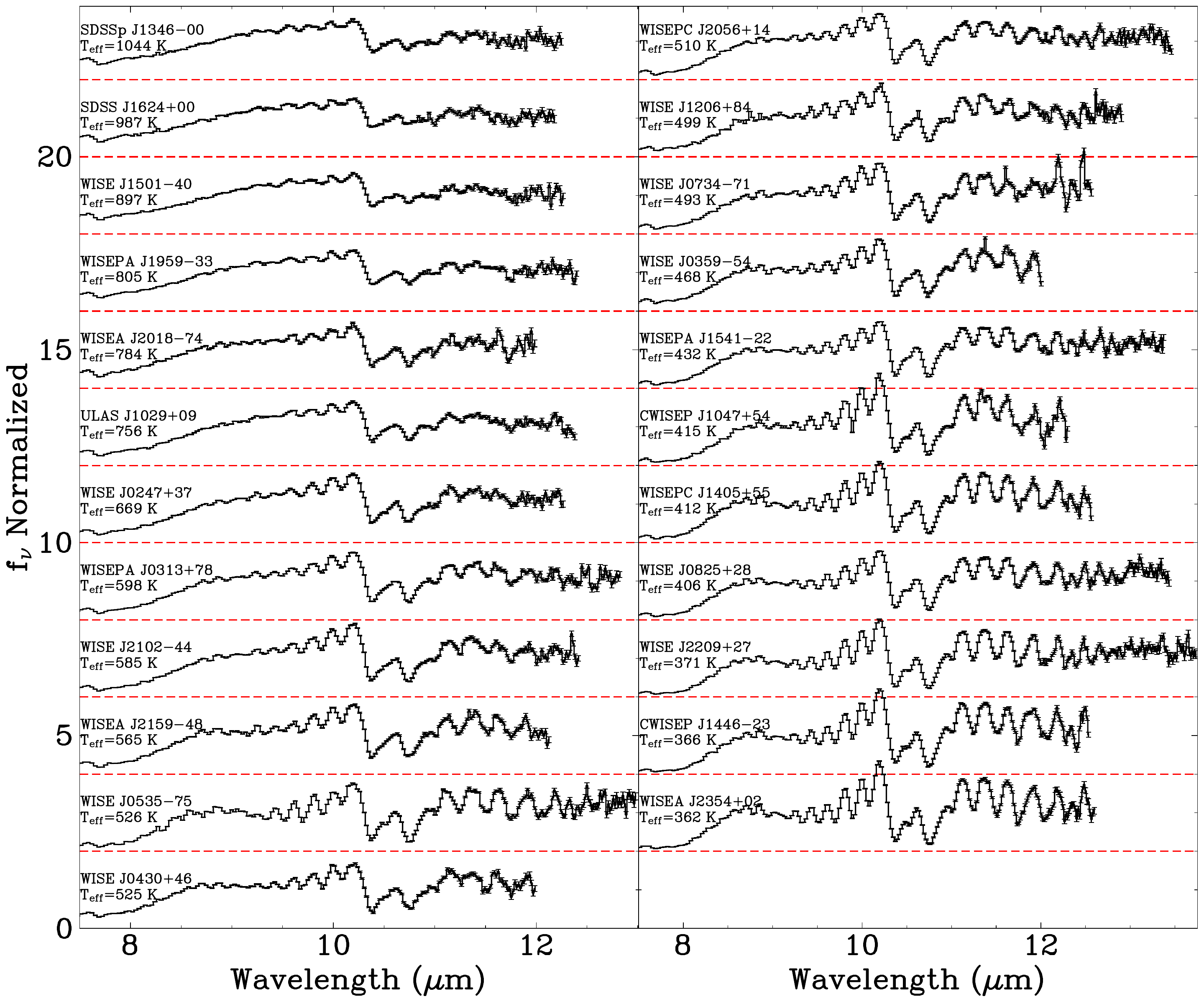}
\centering
\caption{Same as Figure \ref{fig:AllLadderNearIR}, but zoomed in on the 10 \mic{} peak.} \label{fig:AllLadder12mic}
\end{figure*}

We now order our objects by their effective temperatures, as shown across Figures \ref{fig:AllLadderNearIR}$-$\ref{fig:AllLadder12mic}, and observe trends in their spectral morphology. The objects selected for the spectral type sequence in Figure \ref{fig:AbsorptionSequenceAbsNorm} happen to be ordered by effective temperature, but we can see in Figure \ref{fig:AllLadderNearIR} that the near-infrared spectral type ordering does not map directly to effective temperature.  There is still a correlation, as the near-infrared region is primarily shaped by methane and water, whose abundances are strongly dependent on temperature. However there are some objects that buck the spectral type ordering. WISE J0535$-$75 (Y1) is the most discrepant, with a temperature hotter than any other Y dwarf and one T8. This T8, WISE J0430+46, is also out of place. It is colder than any of the T9 dwarfs, and is almost 150 K colder than the other T8 dwarfs. These objects are not poorly spectral typed, as when this sample is ordered by their spectral types, there is a clear sequence in the near-infrared. The problem is that for these cold objects, this region no longer dominates the spectral energy distribution, and as such the near-infrared does not correlate as strongly with \teff{} as it does for hotter objects. 

One would hope that since the 5 \mic{} peak is where a significant portion of our sample's flux is emitted there would be a clear sequence in this wavelength range. However, as seen in Figure \ref{fig:AllLadder5mic}, the 5 \mic{} region does not follow a temperature sequence, nor could we identify one by eye. This lack of a sequence is especially true amongst the Y dwarf. The spectral diversity is due to the same significant atmospheric degeneracies we discussed in regards to the spectral type sequence. Given the spectral diversity and apparent sensitivity of this wavelength range, this spectral region is ripe for further study and will be useful in refining our knowledge of atmospheric physics and chemistry. One prime example of this is how the lack of any phosphine absorption in any of our objects and other published ultracool dwarfs (aside from one reported detection of PH$_3$ in a spectrum with a signal-to-noise ratio of $\sim$5 \citet{burgasser_uncover_2023}) is challenging our current understanding of phosphorus chemistry in these atmospheres.

One possibility for this spectral diversity is that we are probing variation in the surface gravities of these objects. \citet{zahnle_methane_2014} showed how the carbon species are particularly sensitive to gravity and predicted that CO$_2$ could be detectable down to 500 K in low-gravity objects. We do see multiple detections in Y dwarfs below 400 K, which generally have low surface gravities predicted by evolutionary models \citep[e.g.][]{marley_sonora_2021}. The lower gravities of the Y dwarfs also mean that if the variation in gravity remains the same, it would result in a high fractional change in the population, thus creating the greater diversity that we see among the Y dwarfs. Further work will need to be done examining this possible connection between the carbon-bearing species and surface gravity.

There are two objects in particular that stand out due to the features of the carbon-bearing species: CWISEP J1047+54 and WISE J0535$-$75. CWISEP J1047+54 is a Y1 dwarf exhibiting abnormally strong CO$_2$ and CO features compared to other Y1's. Along with this, it also exhibits a steeper slope on the blue edge of the 5 \mic{} peak, likely caused by weaker CH$_4$ absorption. This may be explained by extreme disequilibrium chemistry, low gravity, or a low C/O ratio causing the oxygen bearing molecules to be favored. On the other extreme is WISE J0535$-$75, a Y1 dwarf with no discernible CO$_2$ band and an almost undetectable CO band, possibly due to a low metallicity or high surface gravity. These features result in both objects being distinct in [4.5] vs [3.6]$-$[4.5] color-magnitude space when compared to the rest of the 20 pc sample because of their extreme colors (CWISEP J1047+54 is extremely blue and WISE J0535$-$75 is extremely red). This may mean Spitzer colors are useful for identifying objects with interesting chemistries. 

One major molecule not seen in this wavelength range is NH$_3$, but it can be seen in the NIRSpec portion of the spectrum. A tentative detection of ammonia at 3 \mic{} was claimed in \citet{beiler_first_2023}, and we can confirm it with this full sample, as it is seen in a majority of our objects. The depth of the feature generally increases with temperature, though the low signal-to-noise ratio makes it difficult to determine the depth for the coldest objects.

The ammonia sequence is more apparent in the mid-infrared regime, where the 8$-$12 \mic{} ammonia band absorption increases as effective temperature drops. Only WISE J0535$-$75 breaks the sequence, having stronger absorption than objects with similar temperatures. That said, the spectral morphology of this wavelength range appears to be the most strongly correlated with effective temperature. This region is almost entirely shaped by ammonia, whose abundance is not strongly affected by disequilibrium chemistry, C/O, or M/H \citep{mukherjee_picaso_2023}. Instead, due to being quenched deep in the convective layer, the ammonia abundance and absorption is primarily dependent on effective temperature. This makes the spectral morphology of this region a great tracer of effective temperature.

\section{Conclusion}
We present 0.6$-$22 \mic{} spectral energy distributions of 23 late-T and Y dwarfs, consisting of JWST NIRSpec PRISM, MIRI LRS, and MIRI photometry. With this large sample, we test the absolute calibration of the JWST pipeline by comparing synthetic photometry with previous observations. The NIRSpec data matches most previous observations, though it reveals a systematic offset ($\sim$0.3 mag) between the Spitzer observed and synthetic [3.6] magnitudes that only occurs for late brown dwarfs. This was previously noted in two Y dwarfs, and this large sample confirms that the offset is a systematic issue. We propose that this is caused by the uncommon red shape of the brown dwarf spectrum in this band and/or a light leak.  The MIRI LRS data shows a $\sim$0.04 mag offset when compared to the observed F1000W photometry, which we attribute to the pathloss correction step of the JWST pipeline not accounting for actual slit pointing.

With these broad wavelength spectral energy distributions and precise parallaxes, we are able to calculate the bolometric luminosities of our sample with a precision of $\sim$6\%. From this we calculate \teff{} using three different radii assumptions, including a naive assumption of 1 R$_\mathrm{Jup}$ and two based on evolutionary models and age distributions endorsed by \citet{kirkpatrick_field_2021} and \citet{best_volume-limited_2024}, respectively. This resulted in effective temperatures ranging from $\sim$1000-350 K for spectral types from T6 to Y1. These effective temperatures are approaching the limit of precision achievable with current evolutionary models and age estimates, and extends the population of brown dwarfs with precise effective temperatures into the Y dwarf regime. We also investigate how partial spectral energy distributions systematically affect the derived effective temperatures for these late-type brown dwarfs, and provide corrections as a function of [4.5] absolute magnitude.

Our spectral energy distributions also reveal a plethora of absorption features from a variety of molecules, and with this large sample we can see how these features evolve as a function of spectral type and effective temperature. We note the extreme diversity of the spectral morphology from 3.5$-$5.5 \mic, which appears to be sensitive to atmospheric properties. This includes a surprising lack of PH$_3$, as has been seen in other JWST observations, and strong CO and CO$_2$ absorption even in the Y dwarfs. The diversity of the 5 \mic{} peak is in direct contrast to the 8$-$12 \mic{} regime, whose spectral morphology (shaped largely by NH$_3$) appears to correlate with effective temperature. We also confirm the NH$_3$ detection at 3 \mic{} that was tentatively noted in \citet{beiler_first_2023}. Given that these broad wavelength spectral energy distributions cover a wide range of molecular features, probe various layers of the photosphere, and display significant diversity, this sample is ideal for both forward model and retrieval fits.

\section*{Acknowledgments}
This work is based [in part] on observations made with the NASA/ESA/CSA James Webb Space Telescope. The data were obtained from the Mikulski Archive for Space Telescopes at the Space Telescope Science Institute, which is operated by the Association of Universities for Research in Astronomy, Inc., under NASA contract NAS 5-03127 for JWST. The specific observations analyzed can be accessed via \dataset[10.17909/dwnb-jv28]{https://doi.org/10.17909/dwnb-jv28}. These observations are associated with program \#2302. Support for program \#2302 was provided by NASA through a grant from the Space Telescope Science Institute, which is operated by the Association of Universities for Research in Astronomy, Inc., under NASA contract NAS 5-03127. This research has benefited from the Y Dwarf Compendium maintained by Michael Cushing at \url{https://sites.google.com/view/ydwarfcompendium}. This research has made use of the Spanish Virtual Observatory (https://svo.cab.inta-csic.es) project funded by MCIN/AEI/10.13039/501100011033/ through grant PID2020-112949GB-I00.

\appendix
\section{Synthetic JWST Photometry of Our Sample}\label{sec:app}
We provide synthetic JWST vega magnitudes for our sample of complete spectral energy distributions in all filters that fall entirely within our observed wavelength range in Tables \ref{tbl:SynMags1}-\ref{tbl:SynMags3}. All values were calculated following the process from \citet{gordon_james_2022}. The NIRCam zero points used are the average zero point across all detectors listed in the JWST User Documentation\footnote{\url{https://jwst-docs.stsci.edu/files/182256933/224166043/1/1695068757137/NRC\_ZPs\_1126pmap.txt}}. No table is currently published for the MIRI photometry. Instead we used publicly available Stage 3 data products from JWST Program 1040, which report flux and vega magnitudes, to derive the zero points for filters F560W, F770W, and F1130W (113.69, 64.03, and 23.83 Jy, respectively). The coronagraphic filters do not have any such available observations, so the calculated SVO zero points were used\footnote{\url{http://svo2.cab.inta-csic.es/svo/theory/fps3/}}.

\begin{deluxetable*}{l|c|c|c|c|c|c|c|c|c|c|c|c}[h!]
\tablecaption{Synthetic JWST Photometry of Our Sample\label{tbl:SynMags1}}
\tabletypesize{\footnotesize}
\tablehead{
\colhead{Obj. Name      } &
\colhead{Parallax}&
\colhead{F090W} &
\colhead{F115W} &
\colhead{F140M} &
\colhead{F150W} &
\colhead{F150W2} &
\colhead{F162M} &
\colhead{F164N} &
\colhead{F182M} &
\colhead{F187N} &
\colhead{F200W} &
\colhead{F210M} 
}
\startdata
WISE J0247+37	&	68.4$\pm$2.0&	21.62&	18.63&	20.87&	18.75&	18.74&	18.07&	19.30&	20.07&	20.36&	18.92&	18.13\\
WISEPA J0313+78	&	135.6$\pm$2.8&	20.80&	17.95&	20.37&	18.17&	18.16&	17.48&	18.96&	19.66&	19.81&	18.38&	17.59\\
WISE J0359$-$54	&	73.6$\pm$2.0&	23.17&	21.78&	25.17&	22.07&	21.96&	21.28&	23.06&	23.64&	$\cdots$&	22.17&	21.30\\
WISE J0430+46	&	96.1$\pm$2.9&	21.52&	19.54&	21.69&	19.63&	19.72&	18.90&	19.60&	21.09&	23.03&	20.51&	19.90\\
WISE J0535$-$75	&	68.7$\pm$2.0&	24.11&	22.74&	$\cdots$&	22.76&	22.89&	22.01&	22.43&	23.31&	23.02&	23.08&	22.81\\
WISE J0734$-$71	&	74.5$\pm$1.7&	23.15&	20.73&	23.30&	21.09&	21.01&	20.37&	22.14&	22.49&	22.99&	21.40&	20.62\\
WISE J0825+28	&	152.6$\pm$2.0&	23.19&	22.64&	$\cdots$&	23.34&	22.97&	22.62&	$\cdots$&	23.52&	$\cdots$&	22.55&	21.78\\
ULAS J1029+09	&	68.6$\pm$1.7&	20.56&	17.85&	19.83&	18.01&	17.90&	17.35&	18.34&	18.91&	19.25&	17.81&	17.07\\
CWISEP J1047+54	&	68.1$\pm$4.9&	23.64&	21.45&	26.03&	22.10&	21.87&	21.30&	24.64&	24.18&	24.43&	22.17&	21.28\\
WISE J1206+84	&	84.7$\pm$2.1&	23.41&	20.76&	23.76&	21.27&	21.04&	20.50&	22.36&	22.34&	22.27&	21.09&	20.33\\
SDSSp J1346$-$00	&69.2$\pm$2.3&	18.91&	16.27&	17.79&	16.37&	16.26&	15.72&	16.27&	16.83&	17.05&	16.11&	15.46\\
WISEPC J1405+55	&	158.2$\pm$2.6&	22.30&	21.31&	$\cdots$&	21.66&	21.56&	20.83&	22.58&	23.44&	$\cdots$&	21.96&	21.05\\
CWISEP J1446$-$23	&103.8$\pm$5.0&	24.55&	23.13&	$\cdots$&	23.79&	23.40&	22.88&	24.22&	25.52&	24.11&	23.31&	22.48\\
WISE J1501$-$40	&	72.8$\pm$2.3&	19.36&	16.72&	18.40&	16.90&	16.77&	16.25&	16.83&	17.48&	17.82&	16.68&	16.01\\
WISEPA J1541$-$22	&166.9$\pm$2.0&	22.99&	21.40&	23.70&	22.07&	21.78&	21.41&	$\cdots$&	23.02&	$\cdots$&	21.82&	21.09\\
SDSS J1624+00	&	91.8$\pm$1.2&	18.40&	15.85&	17.47&	15.97&	15.87&	15.30&	15.77&	16.56&	16.91&	15.88&	15.24\\
WISEPA J1959$-$33	&83.9$\pm$2.0&	19.91&	17.19&	19.20&	17.35&	17.23&	16.68&	17.70&	18.15&	18.34&	17.13&	16.39\\
WISEPC J2056+14	&	140.8$\pm$2.0&	22.21&	19.60&	23.07&	20.05&	19.89&	19.26&	20.74&	21.55&	21.04&	20.21&	19.44\\
WISE J2102$-$44	&	92.9$\pm$1.9&	21.35&	18.72&	21.17&	18.95&	18.88&	18.24&	19.69&	20.39&	24.03&	18.99&	18.18\\
WISEA J2159$-$48	&73.9$\pm$2.6&	21.97&	19.44&	21.43&	19.59&	19.64&	18.86&	19.89&	21.33&	22.23&	20.35&	19.58\\
WISE J2209+27	&	161.7$\pm$2.0&	$\cdots$&	$\cdots$&	$\cdots$&	$\cdots$&	$\cdots$&	24.44&	21.80&	$\cdots$&	$\cdots$&	$\cdots$&	22.79\\
WISEA J2354+02	&	130.6$\pm$3.3&	23.60&	22.95&	$\cdots$&	23.43&	23.38&	22.27&	22.78&	24.04&	$\cdots$&	23.32&	22.84\\
WISEA J2018$-$74	&83.2$\pm$1.9&	$\cdots$&	$\cdots$&	$\cdots$&	$\cdots$&	$\cdots$&	$\cdots$&	$\cdots$&	$\cdots$&	$\cdots$&	$\cdots$&	$\cdots$\\
\hline
\enddata
\end{deluxetable*}

\begin{deluxetable*}{l|c|c|c|c|c|c|c|c|c|c|c|c}
\tablecaption{Synthetic JWST Photometry of Our Sample\label{tbl:SynMags2}}
\tabletypesize{\footnotesize}
\tablehead{
\colhead{Obj. Name} &
\colhead{Parallax}&
\colhead{F212N} &
\colhead{F250M} &
\colhead{F277W} &
\colhead{F300M} &
\colhead{F322W2} &
\colhead{F323N} &
\colhead{F335M} &
\colhead{F356W} &
\colhead{F360M} &
\colhead{F405N} &
\colhead{F410M} 
}
\startdata
WISE J0247+37	&	68.4$\pm$2.0&	17.82&	19.91&	18.53&	17.71&	17.19&	20.39&	18.96&	16.63&	16.78&	14.51&	14.65\\
WISEPA J0313+78	&	135.6$\pm$2.8&	17.36&	18.90&	17.48&	16.59&	16.04&	18.49&	17.89&	15.46&	15.60&	13.37&	13.46\\
WISE J0359$-$54	&	73.6$\pm$2.0&	21.43&	21.31&	20.10&	19.23&	18.25&	20.15&	19.68&	17.62&	17.82&	15.49&	15.52\\
WISE J0430+46	&	96.1$\pm$2.9&	19.56&	20.07&	18.84&	17.96&	16.81&	19.58&	18.65&	16.14&	16.29&	14.16&	14.25\\
WISE J0535$-$75	&	68.7$\pm$2.0&	$\cdots$&	22.51&	20.58&	19.61&	18.34&	23.35&	20.54&	17.68&	17.88&	15.50&	15.44\\
WISE J0734$-$71	&	74.5$\pm$1.7&	$\cdots$&	22.03&	20.00&	19.05&	18.24&	21.43&	20.37&	17.63&	17.83&	15.42&	15.47\\
WISE J0825+28	&	152.6$\pm$2.0&	21.65&	21.94&	20.17&	19.19&	18.06&	$\cdots$&	20.20&	17.42&	17.65&	15.17&	15.10\\
ULAS J1029+09	&	68.6$\pm$1.7&	16.79&	18.73&	17.65&	16.90&	16.62&	18.62&	18.14&	16.11&	16.22&	14.12&	14.35\\
CWISEP J1047+54	&	68.1$\pm$4.9&	21.00&	22.87&	21.21&	20.26&	19.32&	$\cdots$&	21.78&	18.74&	19.09&	16.26&	16.42\\
WISE J1206+84	&	84.7$\pm$2.1&	20.00&	21.06&	19.60&	18.68&	17.95&	20.15&	19.80&	17.35&	17.56&	15.18&	15.31\\
SDSSp J1346$-$00	&69.2$\pm$2.3&  $\cdots$&	16.83&	16.10&	15.53&	15.23&	16.85&	16.43&	14.75&	14.83&	12.99&	13.27\\
WISEPC J1405+55	&	158.2$\pm$2.6&  $\cdots$&	22.21&	20.01&	18.86&	17.51&	20.82&	19.69&	16.85&	17.12&	14.57&	14.49\\
CWISEP J1446$-$23	&103.8$\pm$5.0&	22.24&	23.13&	21.48&	20.49&	19.47&	22.50&	22.04&	18.86&	19.21&	16.41&	16.41\\
WISE J1501$-$40	&	72.8$\pm$2.3&	15.73&	17.50&	16.75&	16.15&	15.85&	17.47&	17.05&	15.36&	15.43&	13.58&	13.85\\
WISEPA J1541$-$22	&166.9$\pm$2.0& $\cdots$&	21.48&	19.58&	18.51&	17.56&	19.60&	19.40&	16.93&	17.16&	14.71&	14.68\\
SDSS J1624+00	&	91.8$\pm$1.2&	$\cdots$&	16.53&	15.85&	15.27&	14.92&	16.43&	16.05&	14.42&	14.49&	12.70&	12.90\\
WISEPA J1959$-$33	&83.9$\pm$2.0&  16.13&	17.84&	17.00&	16.35&	15.87&	17.56&	17.26&	15.35&	15.45&	13.37&	13.62\\
WISEPC J2056+14	&	140.8$\pm$2.0&	19.22&	20.08&	18.45&	17.50&	16.77&	19.42&	18.60&	16.16&	16.32&	14.04&	14.13\\
WISE J2102$-$44	&	92.9$\pm$1.9&	$\cdots$&	19.67&	18.35&	17.51&	16.95&	19.75&	18.69&	16.38&	16.53&	14.27&	14.38\\
WISEA J2159$-$48	&73.9$\pm$2.6&  $\cdots$&	20.31&	19.18&	18.33&	17.36&	22.87&	19.52&	16.74&	16.96&	14.59&	14.69\\
WISE J2209+27	&	161.7$\pm$2.0&	$\cdots$&	$\cdots$&	21.22&	19.98&	18.40&	20.19&	21.77&	17.75&	18.12&	15.26&	15.20\\
WISEA J2354+02	&	130.6$\pm$3.3&  21.75&	25.59&	21.30&	20.23&	18.61&	21.73&	22.49&	17.97&	18.38&	15.48&	15.41\\
WISEA J2018$-$74	&83.2$\pm$1.9&	$\cdots$&	$\cdots$&	$\cdots$&	$\cdots$&	$\cdots$&	$\cdots$&	$\cdots$&	$\cdots$&	$\cdots$&	$\cdots$&	$\cdots$\\
\enddata
\end{deluxetable*}

\begin{deluxetable*}{l|c|c|c|c|c|c|c|c|c|c|c|c}
\tablecaption{Synthetic JWST Photometry of Our Sample \label{tbl:SynMags3}}
\tabletypesize{\footnotesize}
\tablehead{
\colhead{Obj. Name} &
\colhead{Parallax}&
\colhead{F430M} &
\colhead{F444W} &
\colhead{F460M} &
\colhead{F466N} &
\colhead{F470N} &
\colhead{F480M} &
\colhead{F560W} &
\colhead{F770W} &
\colhead{F1065C } &
\colhead{F1130W } &
\colhead{F1140C }
}
\startdata
WISE J0247+37	&	68.4$\pm$2.0&	14.44&	14.60&	14.72&	14.56&	14.64&	14.60&	15.23&	14.85&	13.38&	12.61&	12.82\\
WISEPA J0313+78	&	135.6$\pm$2.8&	13.19&	13.33&	13.39&	13.26&	13.31&	13.24&	13.86&	13.39&	11.85&	11.02&	11.23\\
WISE J0359$-$54	&	73.6$\pm$2.0&	15.17&	15.40&	15.60&	15.39&	15.52&	15.34&	16.11&	15.33&	13.86&	12.95&	13.17\\
WISE J0430+46	&	96.1$\pm$2.9&	14.07&	14.24&	14.35&	14.28&	14.30&	14.35&	15.27&	14.64&	13.41&	12.58&	12.80\\
WISE J0535$-$75	&	68.7$\pm$2.0&	14.96&	15.12&	14.94&	14.90&	14.90&	14.90&	16.06&	15.14&	13.61&	12.61&	12.82\\
WISE J0734$-$71	&	74.5$\pm$1.7&	15.11&	15.30&	15.39&	15.25&	15.28&	15.17&	15.95&	15.30&	13.71&	12.80&	13.00\\
WISE J0825+28	&	152.6$\pm$2.0&	14.60&	14.77&	14.75&	14.61&	14.69&	14.52&	15.36&	14.33&	12.62&	11.63&	11.84\\
ULAS J1029+09	&	68.6$\pm$1.7&	14.45&	14.44&	14.82&	14.60&	14.64&	14.55&	14.52&	14.22&	12.76&	12.11&	12.32\\
CWISEP J1047+54	&	68.1$\pm$4.9&	16.21&	16.32&	16.85&	16.49&	16.75&	16.33&	16.90&	16.24&	14.42&	13.39&	13.58\\
WISE J1206+84	&	84.7$\pm$2.1&	15.13&	15.26&	15.81&	15.55&	15.67&	15.27&	15.44&	14.75&	13.12&	12.25&	12.45\\
SDSSp J1346$-$00&	69.2$\pm$2.3&	13.72&	13.56&	14.25&	14.03&	14.06&	13.92&	13.45&	13.26&	12.05&	11.51&	11.73\\
WISEPC J1405+55	&	158.2$\pm$2.6&	13.97&	14.18&	14.08&	14.01&	14.04&	13.98&	15.18&	14.35&	12.77&	11.72&	11.93\\
CWISEP J1446$-$23&	103.8$\pm$5.0&	15.89&	16.04&	16.18&	15.96&	16.09&	15.81&	16.66&	15.71&	13.91&	12.83&	13.04\\
WISE J1501$-$40	&	72.8$\pm$2.3&	14.18&	14.05&	14.51&	14.30&	14.37&	14.27&	14.09&	13.89&	12.66&	12.11&	12.33\\
WISEPA J1541$-$22	&166.9$\pm$2.0&	14.20&	14.34&	14.33&	14.16&	14.25&	14.07&	14.87&	13.88&	12.16&	11.19&	11.41\\
SDSS J1624+00	&	91.8$\pm$1.2&	13.01&	13.05&	13.42&	13.25&	13.27&	13.25&	13.24&	13.04&	11.85&	11.34&	11.55\\
WISEPA J1959$-$33	&83.9$\pm$2.0&	13.77&	13.75&	14.27&	14.12&	14.02&	13.91&	13.75&	13.44&	12.12&	11.52&	11.74\\
WISEPC J2056+14	&	140.8$\pm$2.0&	13.85&	13.99&	14.23&	13.98&	14.16&	13.87&	14.29&	13.59&	11.95&	11.06&	11.27\\
WISE J2102$-$44	&	92.9$\pm$1.9&	14.14&	14.28&	14.42&	14.25&	14.38&	14.23&	14.74&	14.40&	12.87&	11.98&	12.19\\
WISEA J2159$-$48	&73.9$\pm$2.6&	14.44&	14.63&	14.67&	14.61&	14.60&	14.66&	15.61&	15.11&	13.64&	12.80&	13.01\\
WISE J2209+27	&	161.7$\pm$2.0&	14.62&	14.86&	15.03&	14.78&	14.95&	14.67&	15.62&	14.62&	12.90&	11.82&	12.03\\
WISEA J2354+02	&	130.6$\pm$3.3&	14.88&	15.13&	15.16&	15.03&	15.09&	15.01&	16.27&	15.36&	13.68&	12.58&	12.79\\
WISEA J2018$-$74	&83.2$\pm$1.9&	$\cdots$&	$\cdots$&	$\cdots$&	$\cdots$&	$\cdots$&	$\cdots$&	14.34&	13.93&	12.69&	12.02&	12.23\\
\enddata
\end{deluxetable*}


\bibliography{sample631.bib}
\bibliographystyle{aasjournal}

\end{document}